\documentclass[twocolumn,showpacs,preprintnumbers,nofootinbib,prd,superscriptaddress,10pt]{revtex4-1}
\usepackage{graphicx,amssymb,amsmath,amsthm,amsfonts,epsfig}

\usepackage[linktocpage]{hyperref}
\usepackage[usenames]{color}
\usepackage{epstopdf}
\usepackage{aas_macros}

\def\pa{\partial}

\newcommand{\ben}{\begin{enumerate}}
\newcommand{\een}{\end{enumerate}}

\def\be{\begin{equation}}
\def\ee{\end{equation}}
\def\bea{\begin{eqnarray}}
\def\eea{\end{eqnarray}}
\newcommand{\beq}{\begin{eqnarray}}
\newcommand{\eeq}{\end{eqnarray}} 
\newcommand{\ba}{\begin{align}}
\newcommand{\ea}{\end{align}}

\usepackage{tensor}
\usepackage{mathtools}
\usepackage{color}
\usepackage{url}
\newcommand\Lie{\pounds} 

\newcommand{\vp}{\varphi}

\newcommand{\defeq}{\equiv}

\newcommand{\pJ}{\tilde{p}}
\newcommand{\PJ}{\widetilde{P}} 

\newcommand{\uJ}{\tilde{u}}

\newcommand{\rhoJ}{\tilde{\varepsilon}}
\newcommand{\gJ}{\tilde{g}}
\newcommand{\muJ}{\tilde{\mu}} 
\newcommand{\SJ}{\widetilde{S}} 
\newcommand{\sigmaJ}{\tilde{\sigma}} 

\newcommand{\nablaJ}{\widetilde{\nabla}}

\newcommand{\RE}{R_{\ast}}
\newcommand{\xiJ}{\tilde{\xi}} 
\newcommand{\til}{\tilde}
\newcommand{\p}{\partial}
\newcommand{\calY}{{\cal Y}} 
\newcommand{\mueff}{\tilde{\mu}_{\text{eff}}} 
\newcommand{\veleff}{\tilde{v}_{\text{eff}}} 

\begin{document}
\title{Torsional oscillations of neutron stars in scalar-tensor theory of gravity}

\author{Hector O. Silva}
\email{hosilva@phy.olemiss.edu}
\affiliation{Department of Physics and Astronomy, The University of Mississippi, University, MS 38677, USA}

\author{Hajime Sotani}
\email{sotani@yukawa.kyoto-u.ac.jp}
\affiliation{Division of Theoretical Astronomy, National Astronomical Observatory of Japan, 2-21-1 Osawa, Mitaka, Tokyo 181-8588, Japan}

\author{Emanuele Berti} \email{eberti@olemiss.edu}
\affiliation{Department of Physics and Astronomy, The University of
Mississippi, University, MS 38677, USA}

\author{Michael Horbatsch} \email{mhorbats@olemiss.edu}
\affiliation{Department of Physics and Astronomy, The University of
Mississippi, University, MS 38677, USA}
\affiliation{School of Physics and Astronomy,
University of Nottingham, Nottingham, NG7 2RD, UK}

\begin{abstract}
We study torsional oscillations of neutron stars in the scalar-tensor
theory of gravity using the relativistic Cowling approximation. We
compute unperturbed neutron star models adopting realistic equations
of state for the neutron star's core and crust. For scalar-tensor
theories that allow for spontaneous scalarization, the crust thickness
can be significantly smaller than in general relativity. We derive the
perturbation equation describing torsional oscillations in
scalar-tensor theory, and we solve the corresponding eigenvalue
problem to find the oscillation frequencies. The fundamental mode
(overtone) frequencies become smaller (larger) than in general
relativity for scalarized stellar models. Torsional oscillation
frequencies may yield information on the crust microphysics {\em if}
microphysics effects are not degenerate with strong-gravity effects,
such as those due to scalarization. To address this issue, we consider
two different models for the equation of state of the crust and we
look at the effects of electron screening. The effect of scalarization
on torsional oscillation frequencies turns out to be smaller than
uncertainties in the microphysics for all spontaneous scalarization
models allowed by binary pulsar observations. Our study shows that the
observation of quasi-periodic oscillations (QPOs) following giant
flares can be used to constrain the microphysics of neutron star
crusts, whether spontaneous scalarization occurs or not.
\end{abstract}


\date{{\today}}
\maketitle


\section{Introduction}

Observations of quasi-periodic oscillations (QPOs) following giant
flares in soft gamma-ray repeaters
\cite{Israel:2005av,Strohmayer:2005ks,Strohmayer:2006py} suggest a
close coupling between the seismic motion of the crust after a major
quake and the modes of oscillations in a magnetar. The analysis of
X-ray data in SGR 1900+14 \cite{Strohmayer:2005ks} and SGR 1806-20
\cite{Strohmayer:2006py} has unveiled a number of periodicities, with
frequencies that agree reasonably well with the expected torsional (or
toroidal shear) oscillation modes of the neutron star (NS) crust: see
e.g.~\cite{Watts:2011kh} for a review, and \cite{Huppenkothen:2014ufa}
for recent progress in explaining apparent discrepancies between
theoretical models and observations. These observations are very
exciting because they allow us, for the very first time, to test NS
oscillation models.


The foundations of crustal torsional oscillation theory in general
relativity (GR) were laid in a classic paper by Schumaker and Thorne
\cite{Schumaker}. Recent work motivated by QPO observations explored
how torsional oscillation frequencies are affected by various physical
effects, including crustal elasticity \cite{Samuelsson:2006tt},
magnetic fields \cite{Sotani:2006at,Colaiuda:2010pc,Gabler:2011am},
superfluidity \cite{Sotani:2012xd}, the nuclear symmetry energy
\cite{Deibel:2013sia,Iida:2013fra,Sotani:2013jya} and electron
screening \cite{Sotani:2014dua}.

The main motivation of this paper is to answer the following question:
could torsional oscillation frequencies carry observable imprints of
strong-field dynamics, and possibly hint at dynamics beyond GR? Vice
versa, can we ignore effects due to hypothetical strong-field
modifications of GR when we explore the dependence of
torsional oscillation frequencies on the various physical mechanisms
listed above?

We address these questions within the simplest class of modifications
of GR, namely scalar-tensor theory. Damour and Esposito-Far\`ese
\cite{Damour:1993hw} showed that a wide class of scalar-tensor
theories can pass Solar System tests and exhibit nonperturbative
strong-field deviations away from GR (``spontaneous scalarization'')
that can potentially be measured by observations of the bulk
properties of NSs, and of binary systems containing NSs.  The
magnitude of these deviations is very sensitive to the value of a
certain theory parameter $\beta$, defined in Eq.~(\ref{abeta})
below\footnote{There exists a threshold $\beta_c \sim -4.5$, whose
  exact value depends on the NS equation of state.  Scalarization is
  possible when $\beta < \beta_c$.}.

Static NSs in theories with spontaneous
scalarization were first studied in \cite{Damour:1993hw}. 
Their stability was investigated using catastrophe theory by 
Harada 
\cite{Harada:1997mr,Harada:1998ge}.
The
formation of scalarized NSs in gravitational collapse was studied in
\cite{Novak:1998rk,Novak:1999jg}, and a possible mechanism to ``seed''
macroscopic scalar fields from quantum vacuum instabilities was
recently suggested \cite{Lima:2010na,Pani:2010vc,Mendes:2013ija}.
Slowly rotating NSs were studied at first
\cite{Damour:1996ke,Sotani:2012eb} and second \cite{Pani:2014jra}
order in rotation by extending the Hartle-Thorne formalism
\cite{Hartle:1967he,Hartle:1968si}. Recent
work~\cite{Doneva:2013qva,Doneva:2014uma,Doneva:2014faa} addressed the
properties of rapidly rotating NS models.

Widely-separated binary systems of compact objects in scalar-tensor
theory have been studied in
\cite{Damour:1992we,Damour:1996ke,Damour:1998jk}, and the results have
been combined with binary pulsar timing data in order to obtain bounds
on scalar-matter coupling parameters, in particular $\beta$.  Recent
pulsar timing data continue to improve these bounds
\cite{Freire:2012mg,Antoniadis:2013pzd}.
Recently there has been interest in close binaries and mergers, and it
was found that {\it dynamical scalarization} may take place: a close
NS binary may scalarize even if the NSs would not scalarize in
isolation \cite{Barausse:2012da,Shibata:2013pra,Palenzuela:2013hsa}.
The possibility of exploiting this mechanism in order to obtain bounds
on scalar-matter coupling parameters from future gravitational wave
observations has been explored in
\cite{Sampson:2014qqa,Taniguchi:2014fqa}.

A second motivation for this work comes from the surprising finding
that there are universal ``I-Love-Q'' relations between a NS's moment
of inertia, tidal Love number and quadrupole moment in GR
\cite{Yagi:2013bca,Yagi:2013awa}. These relations are ``universal'' in
the sense that they are independent of the poorly known equation of
state (EOS) of matter at high densities. Yagi and Yunes
\cite{Yagi:2013bca,Yagi:2013awa} pointed out that if these relations
were different in alternative theories of gravity, measurements of
these bulk NS properties could be used to constrain alternative
theories or even hint at possible strong-field modifications of
GR. However, stellar structure calculations in scalar-tensor theories
show that the I-Love-Q relations are remarkably insensitive to
scalarization for values of the theory parameters allowed by binary
pulsar tests \cite{Pani:2014jra,Doneva:2014faa}. If the static
properties of NSs (multipole moments and tidal deformation
coefficients) cannot be used for this purpose, it seems natural to
explore QPOs and torsional oscillation frequencies as promising
observational avenues to look for smoking guns of new physics.

Several papers have investigated the signature of alternative theories
of gravity on the NS oscillation spectrum. Sotani et al. studied
nonradial oscillations in scalar-tensor gravity
\cite{Sotani:2004rq,Sotani:2005qx,Sotani:2014tua}, TeVeS \cite{Sotani:2009nm,Sotani:2009xw,Sotani:2010re,Sotani:2011rt} and
Eddington-inspired Born-Infeld gravity \cite{Sotani:2014xoa}. In
particular, Refs.~\cite{Sotani:2004rq,Sotani:2005qx} showed that the
nonradial oscillation frequencies of NSs can change when the effects
of scalarization are large enough to modify the bulk properties of the
star by an appreciable amount. These studies were motivated by
gravitational-wave asteroseismology, i.e. by the prospect of
constraining the stellar properties and the EOS from direct
observations of gravitational radiation from oscillating NSs.  This is
one of the major science goals of third-generation gravitational-wave
detectors such as the Einstein Telescope, but it seems highly unlikely
that we will measure NS oscillation accurately enough to constrain
alternative theories of gravity with upcoming second-generation
experiments, such as Advanced LIGO and Virgo
(cf.~\cite{Andersson:2009yt,Andersson:2013mrx} for reviews).  The
connection between torsional oscillations and QPOs means that our
results have more immediate experimental relevance.

Another noteworthy aspect of this work is that, whereas models of NSs
in alternative theories of gravity usually adopt simple EOS models,
none of these investigations have studied the effect of scalarization
on the structure of the NS crust.  Here we show quantitatively the
connection between the crustal depth, the threshold for scalarization
and the scalar field profile in a scalarized star.

The plan of the paper is as follows.  In Sec.~\ref{sec:background} we
give the equations of hydrostatic equilibrium and we present numerical
results for the equilibrium structure using different models for the
EOS prevailing in the crust. In Sec.~\ref{sec:oscillations} we derive
the perturbation equation describing torsional oscillations in
scalar-tensor theory in the Cowling approximation, and we describe the
numerical method we used to solve the corresponding eigenvalue
problem.  Sec.~\ref{sec:spectra} shows our numerical results for the
oscillation spectra. In the conclusions we discuss the implications
and possible extensions of our work. Appendix~\ref{app1} provides the
derivation of an approximate analytical expression for the ratio
between the crust thickness and the stellar radius in scalar-tensor
theory, that generalizes a similar result by Samuelsson and Andersson
\cite{Samuelsson:2006tt} in GR. We carry out most of the work in the
Einstein frame, but in Appendix~\ref{app2} we show that the Einstein-
or Jordan-frame formulations are equivalent, in the sense that the
energy-momentum conservation law in either frame leads to the same
perturbation equations.

\section{Stellar models in Scalar-Tensor theory}
\label{sec:background}

\subsection{Action and field equations}
We consider the Einstein-frame action \cite{Damour:1993hw} 
\begin{align}
S &= \frac{c^4}{16\pi G_{\ast}}\int d^4x \frac{\sqrt{-g_{\ast}}}{c} \left( R_{\ast} - 2 g_{\ast}^{\mu\nu}\p_{\mu}\vp\p_{\nu}\vp \right) \nonumber \\
&+ S_{\text{M}}\left[\psi_{\text{M}};A^2(\vp)g_{\ast\mu\nu} \right],
\label{action}
\end{align}
where $G_{\ast}$ is the bare gravitational constant, $g_{\ast} \defeq
\text{det}\left[\, g_{\ast\mu\nu}\right]$ is the determinant of the
Einstein-frame metric $g_{\ast\mu\nu}$, $\RE$ is the Ricci curvature
scalar of the metric $g_{\ast\mu\nu}$ and $\varphi$ is a massless scalar
field. $S_{\text{M}}$ is the action of the matter fields
$\psi_{\text{M}}$, coupled to the Einstein-frame metric
$g_{\ast\mu\nu}$ and scalar field $\varphi$ via the Jordan-frame metric
$\tilde{g}_{\mu\nu}\defeq A^2(\varphi)g_{\ast\mu\nu}$, where
$A(\varphi)$ is a conformal factor. 
Throughout this work we use
geometrical units ($c=1=G_{\ast}$) and a mostly plus metric signature
$(-,+,+,+)$. Quantities associated with the Einstein (Jordan) frame
will be labeled with an asterisk (tilde).

The field equations of this theory, obtained by varying the action $S$
with respect to $g^{\mu\nu}_{\ast}$ and $\vp$, respectively, are given
by
\begin{align}
R_{\ast\mu\nu} &= 2 \p_{\mu}\vp\p_{\nu}\vp + 8 \pi \left( T_{\ast\mu\nu} - \frac{1}{2} T_{\ast} g_{\ast\mu\nu}\right), 
\label{field_g} \\
\Box_{\ast} \vp &= -4\pi\alpha(\vp) T_{\ast},
\label{field_phi}
\end{align}
where $R_{\ast\mu\nu}$ is the Ricci tensor, $\alpha(\vp) \defeq d{\log
  A(\vp) }/d\vp$ is usually called the ``scalar-matter coupling
function'', $T_{\ast}^{\mu\nu}$ is the matter field energy-momentum
tensor defined as
\be
T_{\ast}^{\mu\nu} \defeq  \frac{2}{\sqrt{-g_{\ast}}} \frac{\delta S_{\text{M}}\left[ \psi_{\text{M}},A^2(\varphi)g_{\ast\mu\nu} \right]}{\delta g_{\ast\mu\nu}},
\label{t_e}
\ee
and $T_{\ast} \defeq T_{\ast}^{\mu\nu} g_{\ast\mu\nu}$ is its
trace. The energy-momentum tensor in the Jordan frame
$\widetilde{T}^{\mu\nu}$, with trace $\widetilde{T} \defeq
\widetilde{T}^{\mu\nu} \til{g}_{\mu\nu}$, is defined as
\be
\widetilde{T}^{\mu\nu} \defeq  \frac{2}{\sqrt{-\til{g}}} \frac{\delta S_{\text{M}}\left[ \psi_{\text{M}},\til{g}_{\mu\nu} \right]}{\delta \til{g}_{\mu\nu}}.
\label{t_j}
\ee
The energy-momentum tensors (and their traces) in these two
conformally related representations of the theory are related as
follows:
\begin{gather}
T_{\ast}^{\mu\nu} = A^6(\vp) \widetilde{T}^{\mu\nu}, \quad T_{\ast\mu\nu} = A^2(\vp) \widetilde{T}_{\mu\nu}, \nonumber \\ 
T_{\ast} = A^4(\vp) \widetilde{T}.
\label{t_rel}
\end{gather}
Moreover, the covariant divergence of the energy-momentum tensor in
the Einstein and Jordan frames can be shown to be
\begin{align}
\nabla_{\ast_{\mu}} T_{\ast}^{\mu\nu} &= \alpha(\vp) T_{\ast} \nabla_{\ast}^{\nu}\vp,
\label{div_t_e} \\
\widetilde{\nabla}_{\mu} \widetilde{T}^{\mu\nu} &= 0.
\label{div_t_j}
\end{align}
In the limit $\alpha(\vp) \rightarrow 0$ the scalar field decouples
from matter, and the theory reduces to GR.

\subsection{The equations of hydrostatic equilibrium}

The line element describing the space-time of a static, spherically
symmetric star in Schwarzschild coordinates is given by
\be
ds_{\ast}^2 = -e^{2\Phi}dt^2 + e^{2\Lambda}dr^2 + r^2 d\theta^2 + r^2\sin^2{\theta}d\phi^2
\label{metric_e}
\ee
in the Einstein frame, and by
\begin{align}
d\til{s}^2 &= A^2(\vp)\left(-e^{2\Phi}dt^2 + e^{2\Lambda}dr^2 + r^2 d\theta^2 \right. \nonumber \\
&\left. +\, r^2\sin^2{\theta}d\phi^2 \right)
\label{metric_j}
\end{align}
in the Jordan frame, where $\Phi$ and $\Lambda$ are functions of the
radial coordinate $r$. By symmetry, the scalar field $\vp$ also
depends only on $r$.
%
%
We assume the energy-momentum tensor $\tilde{T}_{\mu\nu}$ to be that
of a perfect fluid:
\be
\widetilde{T}_{\mu\nu}=(\rhoJ+\pJ)\uJ_{\mu}\uJ_{\nu}+\pJ\gJ_{\mu\nu},
\label{tmn_fluid}
\ee
where $\rhoJ$ is the energy density, $\pJ$ the pressure and $\uJ_\mu$
the fluid's four-velocity.  Using Eqs.~(\ref{metric_e}) and
(\ref{tmn_fluid}), the field equations (\ref{field_g}) and
(\ref{field_phi}) yield the following equations that describe a static
spherically symmetric star in hydrostatic equilibrium in scalar-tensor
theory \cite{Damour:1993hw,Damour:1996ke}:
\begin{align}
\frac{dm}{dr} &= 4 \pi A^4(\vp) r^2 \rhoJ  + \frac{1}{2}r(r-2m) \psi^2, \label{dm} \\
\frac{d\Phi}{dr} &= 4 \pi A^4(\vp) \frac{r^2 \pJ}{r-2m} + \frac{1}{2}r\psi^2 + \frac{m}{r(r-2m)}, \label{dphi} \\
\frac{d\psi}{dr} &= 4 \pi A^4(\vp) \frac{r}{r-2m}\left[ \alpha(\vp)(\rhoJ - 3\pJ) + r(\rhoJ - \pJ)\psi \right] \nonumber \\ &- \frac{2(r-m)}{r(r-2m)}\psi \label{dpsi}, \\
\frac{d\pJ}{dr} &= - (\rhoJ + \pJ) \left[ \frac{d\Phi}{dr} + \alpha(\vp)\psi \right] \label{dp}.
\end{align}
Here $m=m(r)$ is the relativistic mass-energy function, defined in
terms of $\Lambda(r)$ as $m \defeq (r/2) \left( 1 - e^{-2\Lambda}
\right)$, and we introduced $\psi \defeq d\vp/dr$. 

Hereafter, following Damour and Esposito-Far\`ese
\cite{Damour:1993hw,Damour:1996ke}, we will focus on the scalar-tensor
theory specified by the choice
\be
A(\varphi) = e^{\frac{1}{2} \beta \varphi^2}.
\label{abeta}
\ee
For sufficiently large and negative values of $\beta$, as discussed in
the introduction, NSs in this theory can undergo a phase transition
called \textit{spontaneous scalarization} and acquire a nonvanishing
scalar charge associated with a nontrivial scalar field
configuration. These scalarized solutions of the field equations are
more energetically favorable than non-scalarized solutions.

\begin{figure}[h]
\includegraphics[width=\columnwidth]{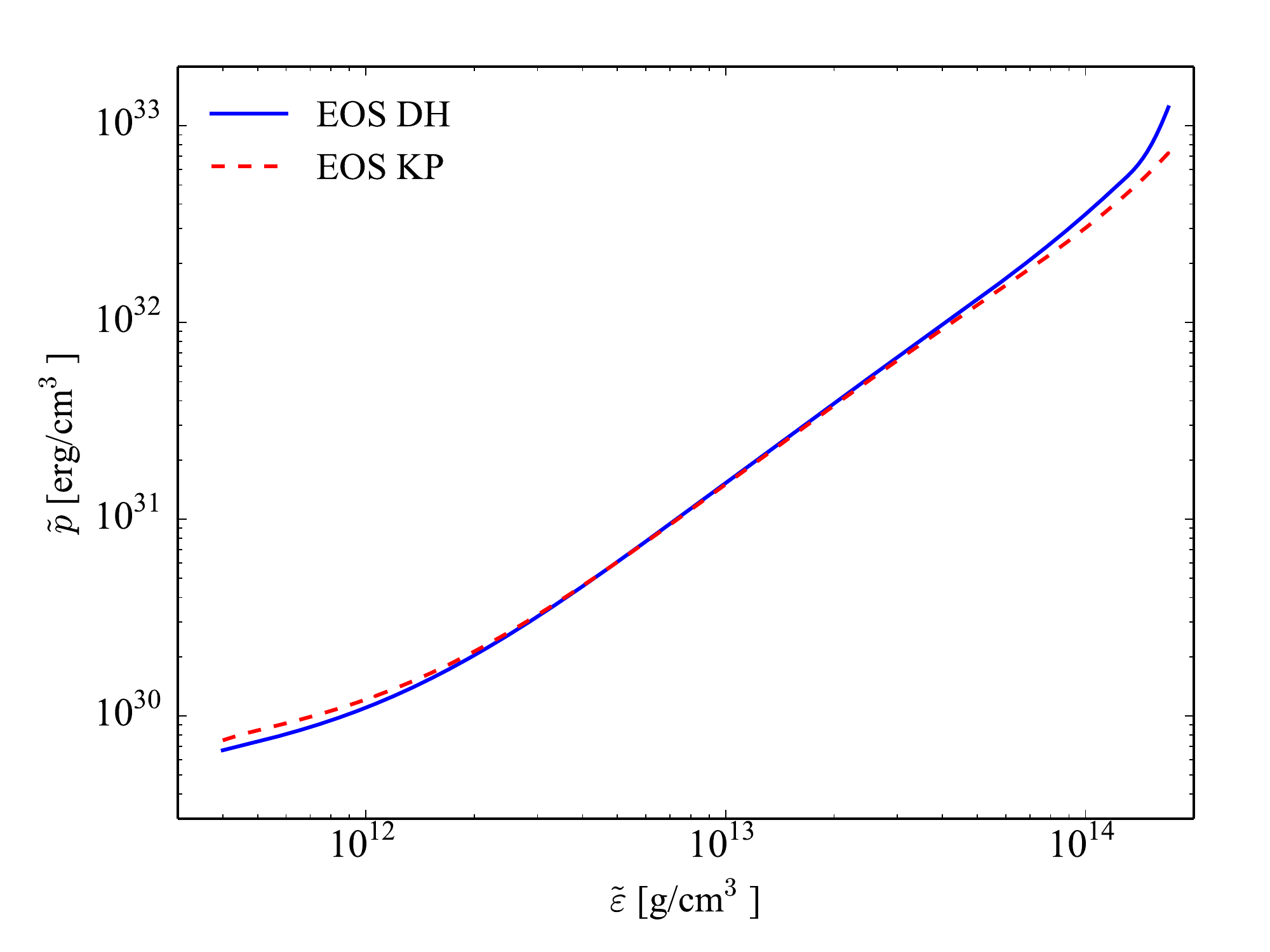}
\caption{(Color online) Pressure $\pJ$ versus energy density $\rhoJ$
  for the crust EOSs considered in this work: EOS DH (solid line) and
  EOS KP (dashed line).}
\label{p_rho}
\end{figure}

To close this system of equations we must complement it with an EOS
$\pJ=\pJ(\rhoJ)$. In this paper, we construct our stellar models
adopting two EOSs for the NS core, namely EOS APR \cite{Akmal:1998cf}
and EOS MS0 \cite{Mueller:1996pm}, while for the NS crust we use the
EOSs derived by Kobyakov and Pethick (henceforth KP,
\cite{Kobyakov:2013eta}) and by Douchin and Haensel (henceforth DH,
\cite{Douchin:2001sv}). These crust EOSs have densities $\rhoJ_b$ at
the crust basis equal to $\rhoJ_b=1.504\times10^{14}$ g$/$cm$^3$ for
EOS KP, and $\rhoJ_b=1.285\times10^{14}$ g$/$cm$^3$ for EOS DH. For a
comparison between the physical assumptions involved in the
construction of these two EOSs, see e.g.~\cite{Sotani:2014dua}. In
Fig.~\ref{p_rho} we display the relation between pressure and energy
density for EOSs DH and KP.

\begin{figure*}[thb]
\includegraphics[width=1.031\columnwidth]{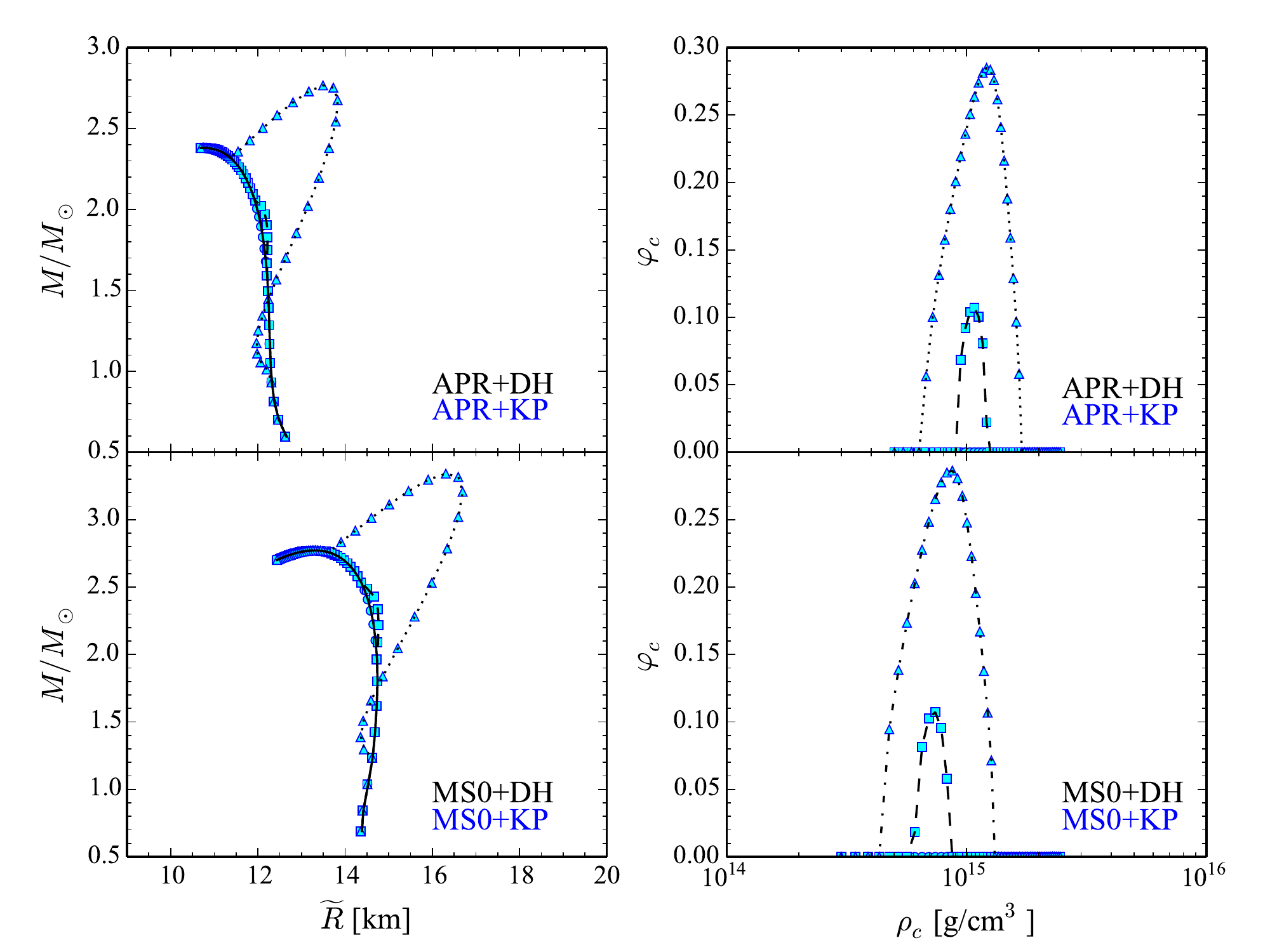}
\includegraphics[width=1.031\columnwidth]{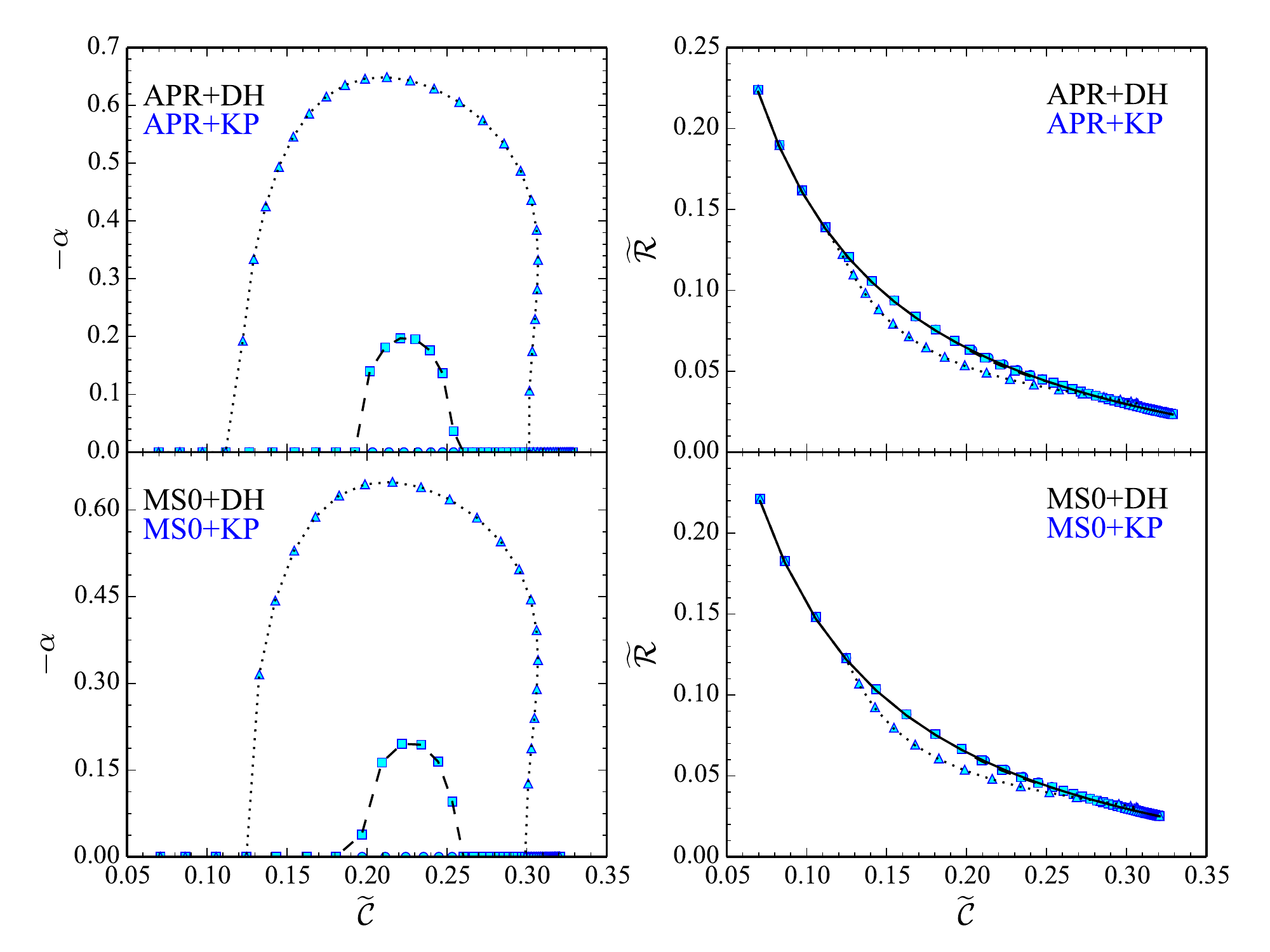}
\caption{(Color online) Properties of our stellar models in
  scalar-tensor theory. From left to right we show: the mass-radius
  relation, the scalar field at the center of the star $\varphi_c$ as a
  function of the central density $\rhoJ_c$, the dimensionless ratio
  $-\alpha=Q/M$ as a function of the compactness $\widetilde{\cal C}$
  and the fractional crust thickness $\widetilde{{\cal R}}$ as a
  function of $\widetilde{{\cal C}}$. The choice of crustal EOS does
  not sensibly affect the crust thickness and the onset of
  scalarization. In all panels, curves with various linestyles
  correspond to stellar models using EOS DH for the NS crust: solid
  lines correspond to $\beta=0.0$, dashed lines to $\beta=-4.5$, and
  dotted lines to $\beta=-6.0$. Different symbols correspond to
  stellar models using EOS KP for the crust: circles for $\beta =
  0.0$, squares for $\beta = -4.5$ and triangles for $\beta = -6.0$.}
\label{props}
\end{figure*}

\subsection{Numerical results for unperturbed stars}

To obtain the equilibrium stellar models we integrate numerically
Eqs.~(\ref{dm})-(\ref{dp}) outwards starting from $r=0$ with initial
conditions $m(0)=0$, $\Phi(0)=\Phi_c$, $\varphi(0)=\varphi_c$,
$\psi(0)=0$ and $\rhoJ(0)=\rhoJ_c$, using one of the two EOSs (APR or
MS0) for the core region. The point $r=r_b$ such that
$\rhoJ(r_b)=\rhoJ_b$ determines the location of the crust basis. The
integration then proceeds until we reach a point $r=r_s$ for which
$\pJ(r_s)=0$, which defines the Einstein-frame radius of the star. The
radii $r_b$ and $r_s$ can be converted to the physical (Jordan) frame
using the relations $\widetilde{R}_b = A(\varphi_b) \, r_b$ and
$\widetilde{R} = A(\varphi_s) \, r_s$, where $\vp_b = \vp(r_b)$ and
$\vp_s = \vp(r_s)$. We can then define the crust thickness as $\Delta
\widetilde{R} \defeq \widetilde{R} - \widetilde{R}_b$. For
convenience, we also introduce the dimensionless fractional crust
thickness $\widetilde{\mathcal{R}} \defeq \Delta \widetilde{R} /
\widetilde{R}$. We remark that the theory is invariant under
reflection symmetry ($\varphi \rightarrow -\varphi$), and therefore,
for simplicity, we shall only consider positive values of the scalar
field.

At spatial infinity ($r \rightarrow \infty$) the metric
$g_{\ast\mu\nu}$ and the scalar field $\varphi$ behave asymptotically
as
\begin{align}
g_{\ast tt} &= -1 + \frac{2 M}{r} +  {\cal O}\left(\frac{1}{r^2}\right),
\label{as-metric-tt} \\
g_{\ast rr} &= 1 + \frac{2 M}{r} +  {\cal O}\left(\frac{1}{r^2}\right),
\label{as-metric-rr} \\
\varphi& = \varphi_{\infty} + \frac{Q}{r}+ {\cal O}\left(\frac{1}{r^2}\right),
\label{as_varphi}
\end{align}
where $M$ is the ADM mass 
and $Q$ is the scalar charge. The values of the various variables at
the stellar surfaces (labeled with the subscript $s$) can be used to
calculate $M$, $Q$ and the asymptotic value of the scalar field
$\varphi_{\infty}\defeq \varphi(r \rightarrow \infty)$ via the
following expressions \cite{Damour:1993hw}:
\begin{align}
M&=r_s^2 \Phi^{\prime}_s \left( 1-\frac{2 m_s}{r_s} \right)^{1/2} \exp\left\{ -\frac{\Phi^{\prime}_s}{\left(\Phi^{\prime 2}_s + \psi^2_s\right)^{1/2}} \right. \nonumber \\
&\left. \times \,  \text{arctanh}\left[ \frac{\left(\Phi^{\prime 2}_s + \psi^2_s\right)^{1/2}}{\Phi^{\prime}_s + 1/r_s} \right] \right\},
\label{adm_m} \\
Q&=-\frac{\psi_s}{\Phi^{\prime}_s} M,
\label{charge}\\
\varphi_{\infty} &= \varphi_s + \frac{\psi_s}{\left(\Phi^{\prime 2}_s + \psi^2_s\right)^{1/2}}\,\text{arctanh}\left[ \frac{\left(\Phi^{\prime 2}_s + \psi^2_s\right)^{1/2}}{\Phi^{\prime}_s + 1/r_s} \right], \nonumber \\
\label{inf_varphi}
\end{align}
where $\Phi^{\prime}_s$ can be calculated with the aid of Eq.~(\ref{dphi}) as
\be
\Phi^{\prime}_s = \frac{1}{2} r_s\, \psi_s^2 + \frac{m_s}{r_s\left( r_s - 2 m_s \right)},
\ee
and primes indicate partial derivatives with respect to the radial
coordinate $r$. From now on, we will assume that $\varphi_{\infty}=0$.

To obtain solutions of Eqs.~(\ref{dm})-(\ref{dp}) satisfying this
assumption, we apply the shooting method in order to find the central
values of the scalar field $\varphi_c$ such that the required value of
$\varphi_{\infty}$ is obtained. As a check of our code we compared our
results against the ones presented in Refs.~\cite{Doneva:2014uma} (in
scalar-tensor theory) and \cite{Kokkotas:2000up,Berti:2004ny} (in GR),
finding excellent agreement.

In Fig.~\ref{props} we present general properties of stellar models
constructed by solving Eqs.~(\ref{dm})-({\ref{dp}}) combining EOS APR
and EOS MS0 (for the NS core) with EOS KP and EOS DH (for the NS
crust). The top row refers to the APR EOS, and the bottom row refers to the
MS0 EOS; results for different crust models are shown using different
linestyles in each inset.

The leftmost column shows the mass-radius relation. Deviations from GR
due to spontaneous scalarization are clearly visible; we also see that
the choice of crustal EOS has negligible influence on the mass-radius
relation, for both ``ordinary'' and scalarized stars. The second
column shows the central value of the scalar field $\varphi_c$ as a
function of the central density $\rhoJ_c$. The scalar field at the
center acquires a nonzero value (i.e., the NS becomes scalarized)
around $\rhoJ_c \approx 4\times10^{14}$ $-$ $6\times10^{14}$
g$/$cm$^3$, and it has a maximum around $\rhoJ_c \approx
7\times10^{14}$ $-$ $9\times10^{14}$ g$/$cm$^3$.
In the third column we plot the dimensionless scalar charge
$\alpha\defeq - Q/M$ as a function of the compactness $\widetilde{\cal
  C} \defeq M/\widetilde{R}$ (both expressed in geometrical
units). Finally, the rightmost column shows $\widetilde{\cal R}$ as a
function of the compactness $\widetilde{\cal C}$.  In comparison with
their GR counterparts, for scalarized stars the crust represents a
smaller fraction of the NS interior. Note also that deviations in the
crust thickness due to scalarization and nonzero scalar charges
develop in the same range of compactness $\widetilde{\cal C}$, as
expected.

These plots show that the choice of crustal EOS has negligible effects
on the bulk properties of the star. This is not surprising,
considering that EOSs DH and KP have very similar crust basis
densities $\rhoJ_b$ and $\pJ(\rhoJ)$ (cf. Fig.~\ref{p_rho}). However,
as we will see in Sec.~\ref{shear}, different crustal EOSs result in
rather different elastic properties for the crust, and they do have an
effect on torsional oscillation frequencies.

\subsection{An approximate formula for ${\cal R}$}

Samuelsson and Andersson~\cite{Samuelsson:2006tt} obtained a simple
approximate analytical expression for the ratio between the crust
thickness and stellar radius ${\cal R}$, within GR, in
terms of the star's compactness ${\cal C}$:
\be
{\cal R} = \left( \frac{{\cal C}}{\sigma} e^{2 \Lambda} + 1 \right)^{-1},
\label{approx_gr}
\ee
where $e^{-2 \Lambda} = 1 - 2\,{\cal C}$ and $\sigma \approx 0.02326$
is a constant found by curve fitting, which in general depends on the
crustal EOS \cite{Samuelsson-thesis}.

\begin{figure}
\includegraphics[width=\columnwidth]{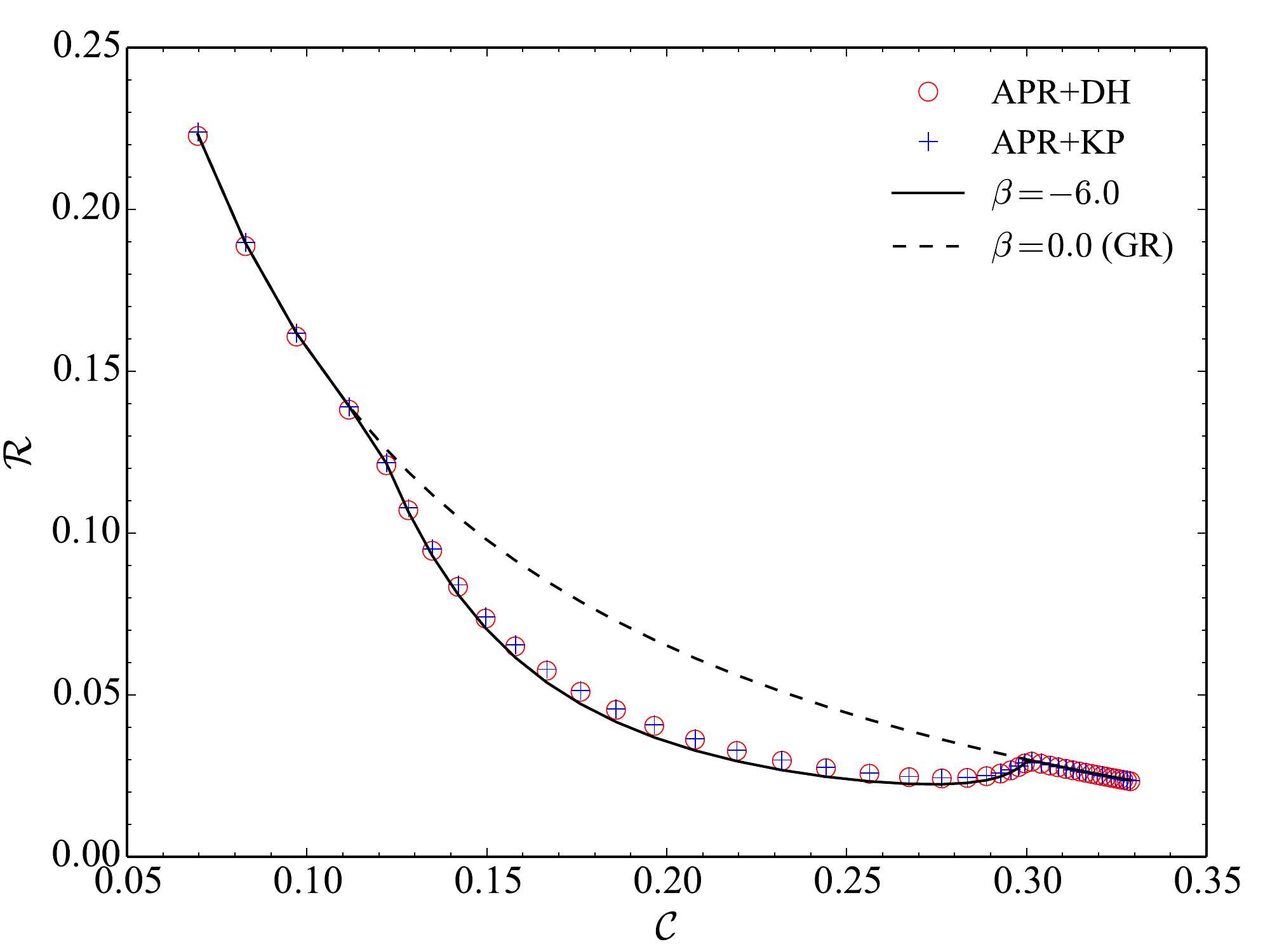}
\caption{(Color online) Comparison between Eq.~(\ref{approx_st}) and
  the numerical results for $\beta = -6.0$, using $\sigma =
  0.02326$. The GR expression (\ref{approx_gr}) is also shown. Since
  the integration of Eqs.~(\ref{dm})-(\ref{dp}), gives us $\vp$ in the
  Einstein-frame radial coordinate $r$, the compactness and fractional
  crust thickness are evaluated in this frame. Notice, however, that
  even for $\beta = -4.5$ (a value marginally excluded by binary
  pulsars observations \cite{Freire:2012mg}) the percent difference
  between the compactnesses and fractional crust thicknesses in the
  two frames is less than $1.0\%$, and therefore Eq.~(\ref{approx_st})
  is accurate for all physically sensible values of $\beta$.}
\label{approx_crust}
\end{figure}

In Appendix~\ref{app1} we show that this result can be generalized to
scalar-tensor theory as follows:
\be
{\cal R} = \frac{\sigma}{ 2 \beta \zeta}\left( {\cal F} - \sqrt{{\cal F}^2-\frac{4 \beta \zeta}{\sigma}}\,\right),
\label{approx_st}
\ee
where we introduced
\be
{\cal F} \defeq 1 + \frac{1}{\sigma} \left( {\cal C} e^{2\Lambda} + \beta \zeta \right)
\ee
and $\zeta = \zeta({\cal C}) \defeq \varphi_s\, \psi_s\, r_s$, which
is obtained by interpolation, given a family of stellar models, as a
function of ${\cal C}$. We make the same approximations used in
\cite{Samuelsson:2006tt}, and in addition we assume the scalar field
to be constant throughout the NS crust. From Eq.~(\ref{approx_st}) we
can also calculate the first correction to Eq.~(\ref{approx_gr}) in
powers of $\beta\zeta$, due the presence of the scalar field in a
scalarized NS:
\begin{align}
{\cal R} \approx \left( \frac{{\cal C}}{\sigma} e^{2 \Lambda} + 1 \right)^{-1} - 2\,{\cal C} e^{2\Lambda}\frac{(\beta \zeta)^2}{\sigma^3} \left( \frac{{\cal C}}{\sigma} e^{2 \Lambda} + 1 \right)^{-3}, \nonumber \\
\label{first_correc}
\end{align}
where the minus sign indicates that ${\cal R}$ is smaller for such
stars in comparison to nonscalarized ones, as observed in
Figs.~\ref{props} and \ref{approx_crust}.

To illustrate how accurately Eq.~(\ref{approx_st}) describes the
behavior of ${\cal R}$ observed in Fig.~\ref{props}, in
Fig.~\ref{approx_crust} we plot ${\cal R}$, choosing EOS APR to
describe the NS core, as a function of ${\cal C}$ for $\beta=-6.0$
(the case in which deviations from GR are greatest). We find good
agreement between the approximate expression and data obtained by
numerically solving Eqs.~(\ref{dm})-(\ref{dp}). As can be seen, the
same value of $\sigma$ obtained in \cite{Samuelsson:2006tt} for the
EOS used in \cite{Samuelsson-thesis} is accurate enough for both EOS
DH and EOS KP.

\section{Torsional perturbations in the Cowling approximation}
\label{sec:oscillations}

\subsection{Derivation of the perturbation equations}
Let us now derive the equation describing torsional oscillations in
scalar-tensor theory. We begin by introducing a small fluid
perturbation described by a Lagrangian displacement vector
\be
\xiJ^{i} = \left( 0,0, \widetilde{\calY}(t,r) \frac{1}{\sin\theta} \p_{\theta}P_{\ell}(\cos\theta) \right),
\label{pert_xi}
\ee
where $P_\ell(\cos\theta)$ is the Legendre polynomial of order
$\ell$. For notational convenience, in Eq.~(\ref{pert_xi}) we omit the
sum over $\ell$. The perturbation of the fluid four-velocity $\delta
\uJ^{3} = \uJ^{0} ( \p \xiJ^{3} / \p t )$ is
\be
\delta \uJ^{3} = A^{-1}(\vp) e^{-\Phi} \dot{\widetilde{\calY}}(t,r)\frac{1}{\sin \theta} \p_{\theta}P_{\ell}(\cos\theta),
\ee
where the dot represents a partial derivative with respect to the time
coordinate $t$.

In this work we use the Cowling approximation
\cite{1983ApJ...268..837M,1988MNRAS.232..259F}, i.e. we assume that
matter perturbations do not result in perturbations on the metric
$\tilde{g}_{\alpha\beta}$: $\delta \gJ_{\mu\nu} = 0$. Within this
approximation, the perturbed perfect fluid energy-momentum tensor
(\ref{tmn_fluid}), including the shear tensor contribution $
\delta\SJ_{\mu\nu}$, is
\be
\delta\widetilde{T}_{\mu\nu} = \left( \pJ + \rhoJ \right) \left( \delta\uJ_{\mu} \uJ_{\nu} + \uJ_{\mu}\delta\uJ_{\nu} \right) - 2 \muJ \delta\widetilde{S}_{\mu\nu},
\label{pert_t}
\ee
as the pressure $\pJ$, the energy density $\rhoJ$ and the scalar field
$\vp$ are unaffected by odd (axial) perturbations. We have also
introduced the shear modulus $\muJ=\muJ(r)$. While the first term in
Eq.~(\ref{pert_t}) is simple to calculate, to obtain
$\delta\SJ_{\mu\nu}$ we must first use the fact that
$\delta\sigmaJ_{\mu\nu} \defeq \Lie_{\uJ} \delta\widetilde{S}_{\mu\nu}
= A^{-1}(\vp) \exp(-\Phi)\partial_0 \delta\widetilde{S}_{\mu\nu}$,
where the perturbed rate of shear
$\delta\sigmaJ_{\mu\nu}=\delta\sigmaJ_{\nu\mu}$ is given by
\begin{align}
\delta\sigmaJ_{\mu\nu} &= \frac{1}{2}\left( \delta \tensor{\PJ}{^{\alpha}_{\nu}} \nablaJ_{\alpha} \uJ_{\mu} + \delta \tensor{\PJ}{^{\alpha}_{\mu}} \nablaJ_{\alpha} \uJ_{\nu} + \tensor{\PJ}{^{\alpha}_{\nu}} \nablaJ_{\alpha} \delta \uJ_{\mu} \right. \nonumber \\
&\left.+ \tensor{\PJ}{^{\alpha}_{\mu}} \nablaJ_{\alpha} \delta\uJ_{\nu} \right) 
- \frac{1}{3} \left( \delta\PJ_{\mu\nu} \nablaJ_{\alpha} \uJ^{\alpha} + \PJ_{\mu\nu} \nablaJ_{\alpha} \delta\uJ^{\alpha} \right), \nonumber \\
\label{pert_sigma}
\end{align}
$\delta \PJ_{\mu\nu}$ denotes the perturbed projection operator
\be
\delta \PJ_{\mu\nu} =  \delta\uJ_{\mu} \uJ_{\nu} + \uJ_{\mu}\delta\uJ_{\nu},
\label{pert_p}
\ee
and $\Lie_{\uJ}$ is the Lie derivative along the worldline of a fluid
element \cite{Schumaker}. The nonzero components of the perturbed rate
of shear $\delta\sigmaJ_{\mu\nu}$ can then be shown to be
\begin{align}
\delta\sigmaJ_{13} &= \frac{1}{2} A(\vp) e^{-\Phi} \dot{\widetilde{\calY}}^{\prime}(t,r) \, r^2 \sin\theta\, \partial_\theta P_{\ell} (\cos\theta), 
\label{pert_sig31}\\
\delta\sigmaJ_{23} &= \frac{1}{2} A(\vp) e^{-\Phi} \dot{\widetilde{\calY}} (t,r) r^2 \sin^2\theta \, \partial_{\theta} \left[ \frac{1}{\sin\theta} \partial_\theta P_{\ell}(\cos\theta) \right]. \nonumber \\
\label{pert_sig32}
\end{align}
Using Eqs.~(\ref{pert_sig31}) and (\ref{pert_sig32}), the perturbed
shear tensor has components
\begin{align}
\delta\SJ_{13} &=  \frac{1}{2} A^2(\vp) \widetilde{{\calY}}^{\prime}(t,r) r^2 \sin\theta\, \partial_\theta P_{\ell} (\cos\theta), \\
\delta\SJ_{23} &= \frac{1}{2} A^2(\vp) \widetilde{{\calY}}(t,r) r^2 \sin^2\theta\, \partial_{\theta} \left[ \frac{1}{\sin\theta} \partial_\theta P_{\ell}(\cos\theta) \right]. \nonumber \\
\end{align}

Combining these results, we find that the nonzero components of the
perturbed energy-momentum tensor are
\begin{align}
\delta\widetilde{T}_{03} &= - (\pJ + \rhoJ) A^{2}(\vp) \dot{\widetilde{\calY}} \,r^2 \sin\theta\, \partial_{\theta} P_{\ell} (\cos\theta), 
\label{pert_t30}
\\
\delta\widetilde{T}_{13} &= - \muJ A^2(\vp) \widetilde{{\calY}}^{\prime}(t,r) r^2 \sin\theta\, \partial_\theta P_{\ell} (\cos\theta),
\label{pert_t31}
\\
\delta\widetilde{T}_{23} &= -\muJ A^2 (\vp) \widetilde{{\calY}}(t,r) r^2 \sin^2\theta\, \partial_{\theta} \left[ \frac{1}{\sin\theta} \partial_\theta P_{\ell}(\cos\theta) \right]. \nonumber\\
\label{pert_t32}
\end{align}
In the GR limit -- obtained by taking $A(\vp)=1$, and consequently
$\alpha(\vp) = 0$ -- the above results are in agreement with
\cite{Schumaker} when we neglect metric perturbations in their
equations.

In the Cowling approximation, the variation of the energy-momentum
conservation law in the Jordan frame \cite{Sotani:2004rq} can be
obtained from Eq.~(\ref{div_t_j})
\begin{align}
\nablaJ_{\nu} \delta{\tensor{\widetilde{T}}{^{\nu}_{\mu}}} &= \partial_{\nu} \delta{\tensor{\widetilde{T}}{^{\nu}_{\mu}}} + \Gamma^{\nu}_{\ast\alpha\nu} \delta{\tensor{\widetilde{T}}{^{\alpha}_{\mu}}} - \Gamma^{\alpha}_{\ast\mu\nu} \delta{\tensor{\widetilde{T}}{^{\nu}_{\alpha}}} \nonumber \\
&+4 \alpha(\vp) \partial_{\alpha}\vp\, \delta{\tensor{\widetilde{T}}{^{\alpha}_{\mu}}} - \alpha(\vp) \partial_{\mu}\vp\, \delta{\tensor{\widetilde{T}}{^{\alpha}_{\alpha}}} \nonumber \\ &= 0,
\label{pert_j}
\end{align}
where $\Gamma^{\mu}_{\ast\nu\sigma}$ denotes the Christoffel symbols
of the Einstein-frame metric, related to they Jordan frame
counterparts by
\begin{align}
\widetilde{\Gamma}^{\sigma}_{\mu\nu} = \Gamma^{\sigma}_{\ast\mu\nu}+\alpha(\vp)\left( \delta^{\sigma}_{\nu} \partial_{\mu}\vp + \delta^{\sigma}_{\mu} \partial_{\nu}\vp - g^{\sigma \rho}_{\ast} g_{\ast\mu\nu} \partial_{\rho} \vp \right). \nonumber \\
\end{align}
In Appendix~\ref{app2} we show that Eq.~(\ref{pert_j}) can also be
obtained starting from the energy-momentum conservation law
(\ref{div_t_e}) in the Einstein frame, and therefore the two frames
are physically equivalent.

By setting $\mu = 3$ and making use of
Eqs.~(\ref{pert_t30})-(\ref{pert_t32}) we obtain the following
differential equation for $\widetilde{\calY}(t,r)$:
\begin{align}
\widetilde{\calY}^{\prime\prime}(r) &+ \left[ \frac{4}{r} + \Phi^{\prime} - \Lambda^{\prime} + \frac{\muJ^{\prime}}{\muJ} + 4 \alpha(\vp) \psi \right]\, \widetilde{\calY}^{\prime}(r) \nonumber \\
&+ \left[\left( \frac{\omega}{\tilde{v}_s}\right)^2e^{-2\Phi} - \frac{(\ell+2)(\ell-1)}{r^2} \right] {e^{2\Lambda}}  \widetilde{\calY}(r) =0, \nonumber \\
\label{eq_y}
\end{align}
where we have assumed a harmonic time dependence $\tilde{\calY}(t,r) =
\tilde{\calY}(r) e^{i \omega t}$ for the perturbation variable, and we
have introduced the shear wave velocity $\tilde{v}_s^2 \defeq \muJ /
(\pJ+\rhoJ)$.

We can recast Eq.~(\ref{eq_y}) in a form identical to the GR case
(cf.~\cite{Schumaker,Sotani:2014dua}) if we introduce an effective
shear modulus $\mueff \defeq A^4(\vp) \muJ$, an effective wave
velocity $\veleff^2 \defeq A^4(\vp) \tilde{v}_s^2$ and a rescaled frequency
$\bar{\omega} = A^2(\vp)\, \omega$:
\begin{align}
\widetilde{\calY}^{\prime\prime}(r) &+ \left[ \frac{4}{r} + \Phi^{\prime} - \Lambda^{\prime} + \frac{\mueff^{\prime}}{\mueff}\right]\, \widetilde{\calY}^{\prime}(r) \nonumber \\
&+ \left[\left( \frac{\bar{\omega}} {\veleff}\right)^2e^{-2\Phi} - \frac{(\ell+2)(\ell-1)}{r^2} \right] {e^{2\Lambda}}  \widetilde{\calY}(r) =0. \nonumber \\
\label{eq_y_2}
\end{align}
Given the definition of the conformal factor (\ref{abeta}), the factor
$A^4(\vp)$ is always less than unity when $\beta<0$, and therefore
$\mueff / \muJ \leq 1$.

To obtain the oscillation frequencies we must integrate
Eq.~(\ref{eq_y_2}) numerically with appropriate boundary
conditions. We assume that torsional oscillations are confined to the
NS crust, so our boundary conditions are a zero-torque condition at $r
= r_s$ and a zero-traction condition at $r_b$. These boundary
conditions follow from the fact that the shear modulus is zero in the
NS core and outside the star, and they imply that
$\widetilde{\calY}(r)$ must satisfy Neumann boundary conditions,
i.e. $\widetilde{\calY}^{\prime}(r) = 0$ at both $r = r_b$ and $r=
r_s$ \cite{Schumaker,Sotani:2014dua,Sotani:2006at}. Our integrations
of Eq.~(\ref{eq_y_2}) are performed in the Einstein frame, but since
$\vp_{\infty} = 0$, the torsional oscillation frequencies measured at
infinity are the same in the Einstein and Jordan frames.

Following common practice in the literature, we will present numerical
results for the torsional oscillation frequencies ${_n}t_{\ell} \equiv
\omega / (2 \pi)$. Here $n$ is the number of radial nodes of the
function $\widetilde{\calY}(r)$ in the crust region, and $\ell$ is the
usual angular index associated with the Legendre polynomials
$P_\ell(\cos\theta)$.

\begin{figure}
\includegraphics[width=\columnwidth]{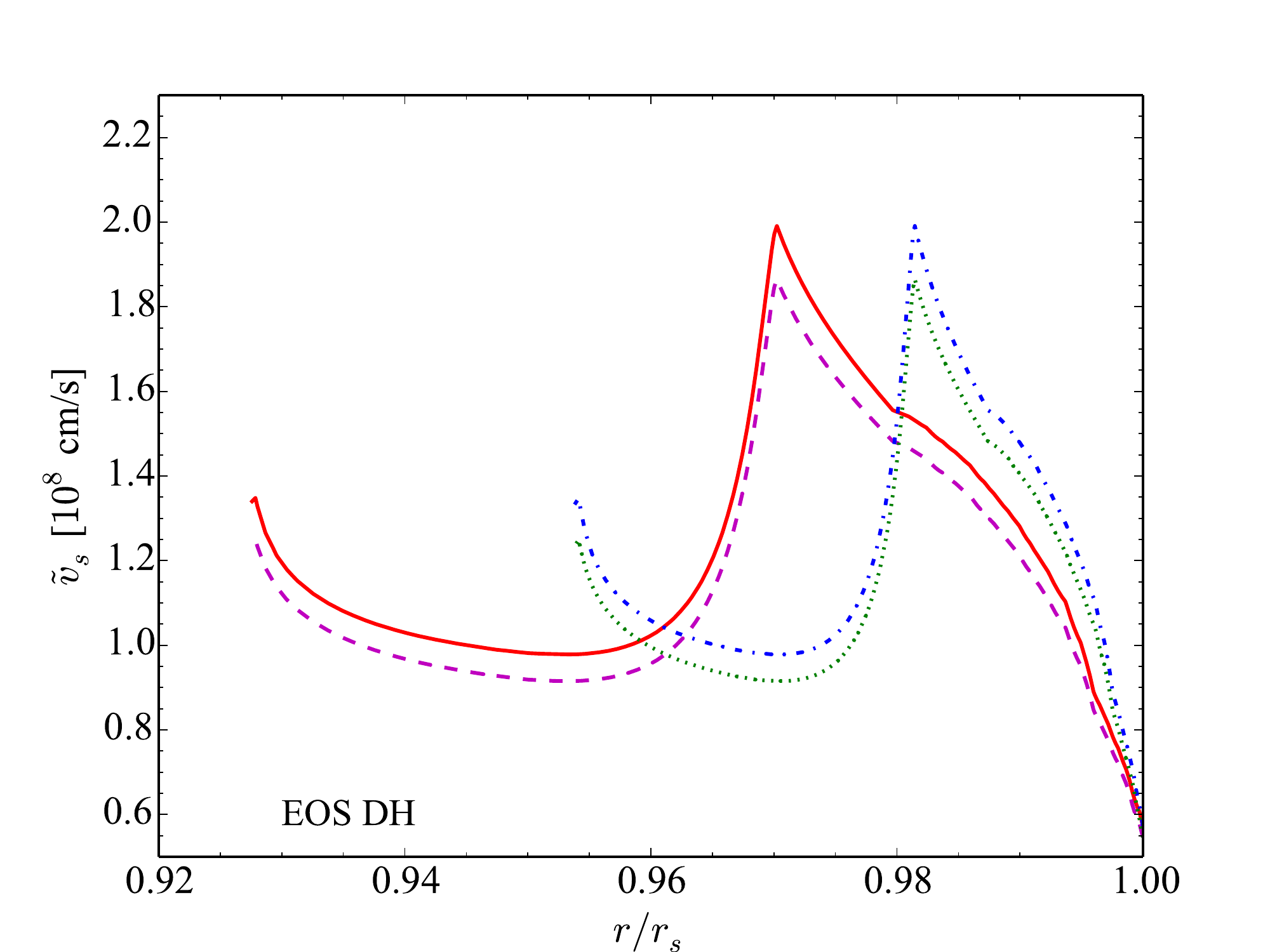}
\includegraphics[width=\columnwidth]{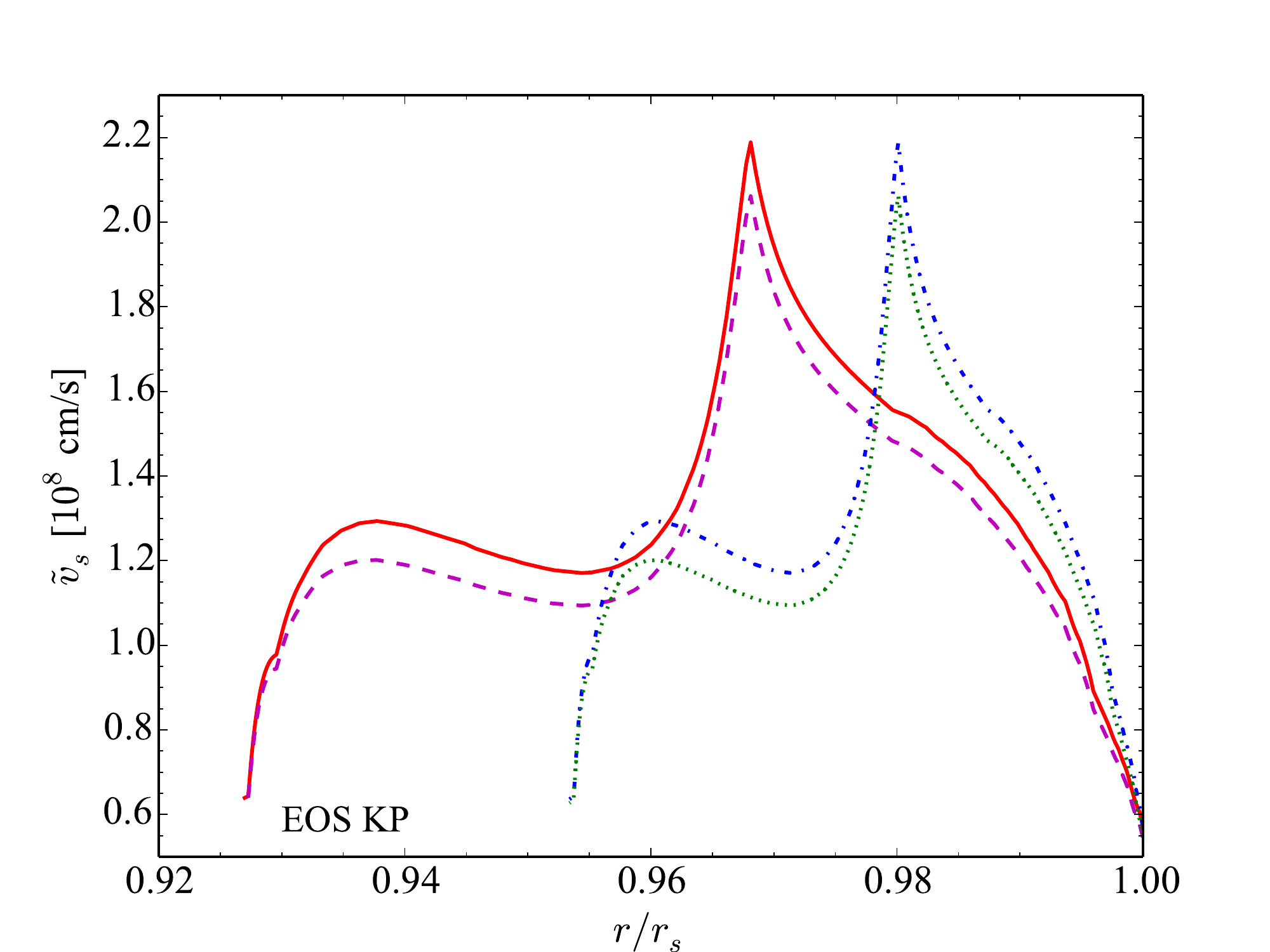}
\caption{(Color online) Shear velocity profile $\tilde{v}_s(r)$ in the
  NS crust in the following cases: (i) GR without electron screening
  (solid line); (ii) GR with electron screening (dashed line); (iii)
  scalar-tensor theory ($\beta = - 6.0$) without electron screening
  (dashed-dotted line); (iv) scalar-tensor theory ($\beta = - 6.0$)
  with electron screening (dotted line). The top panel refers to EOS
  DH, the bottom panel to EOS KP. The sharp peaks occur near the
  neutron drip density $\rhoJ \approx 3 \times 10^{11}$ g/cm$^3$
  \cite{Shapiro:1983du}.}
\label{v_s}
\end{figure}
%

%
\begin{figure*}[thb]
\includegraphics[width=0.95\columnwidth]{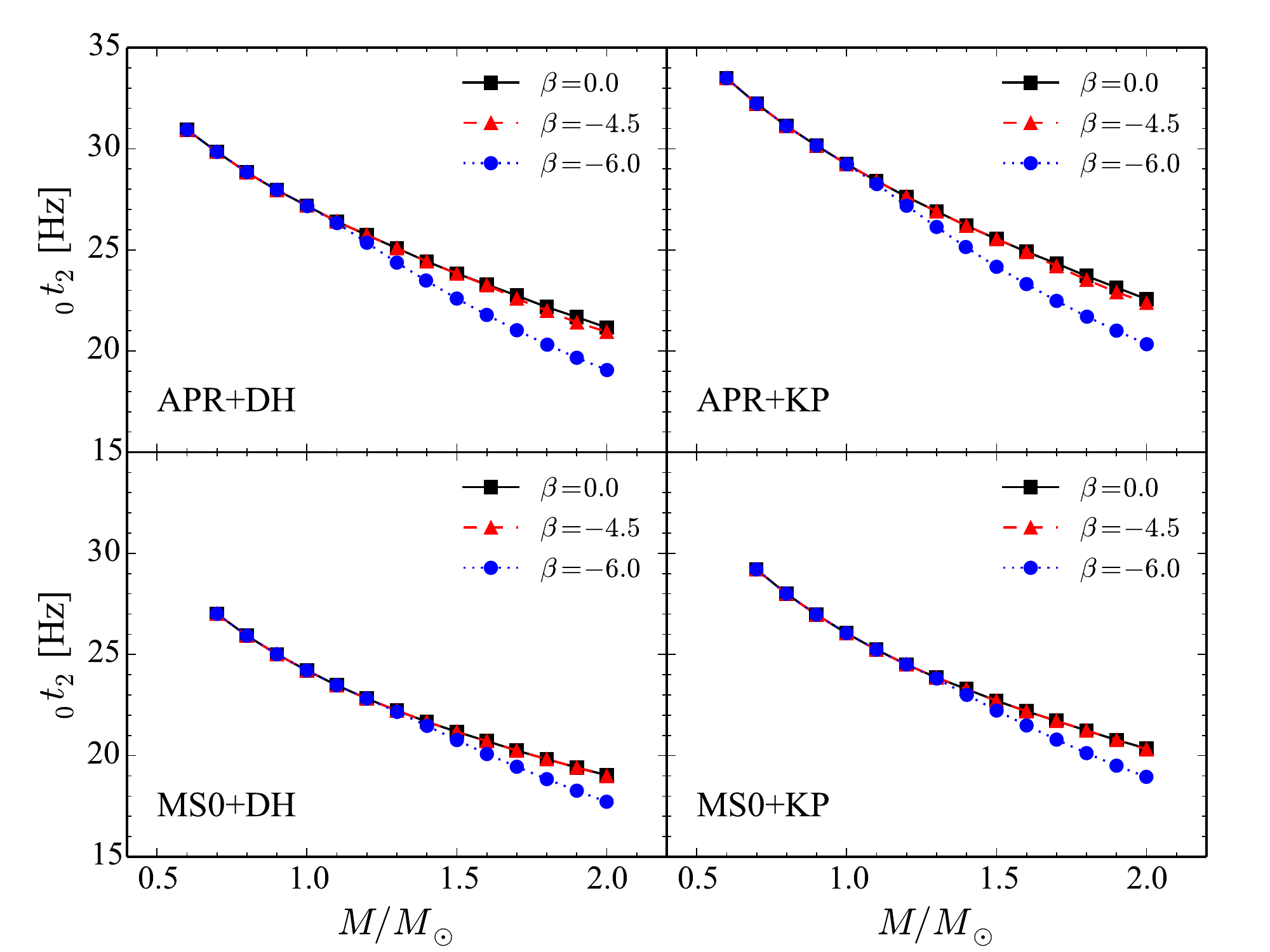}
\includegraphics[width=0.95\columnwidth]{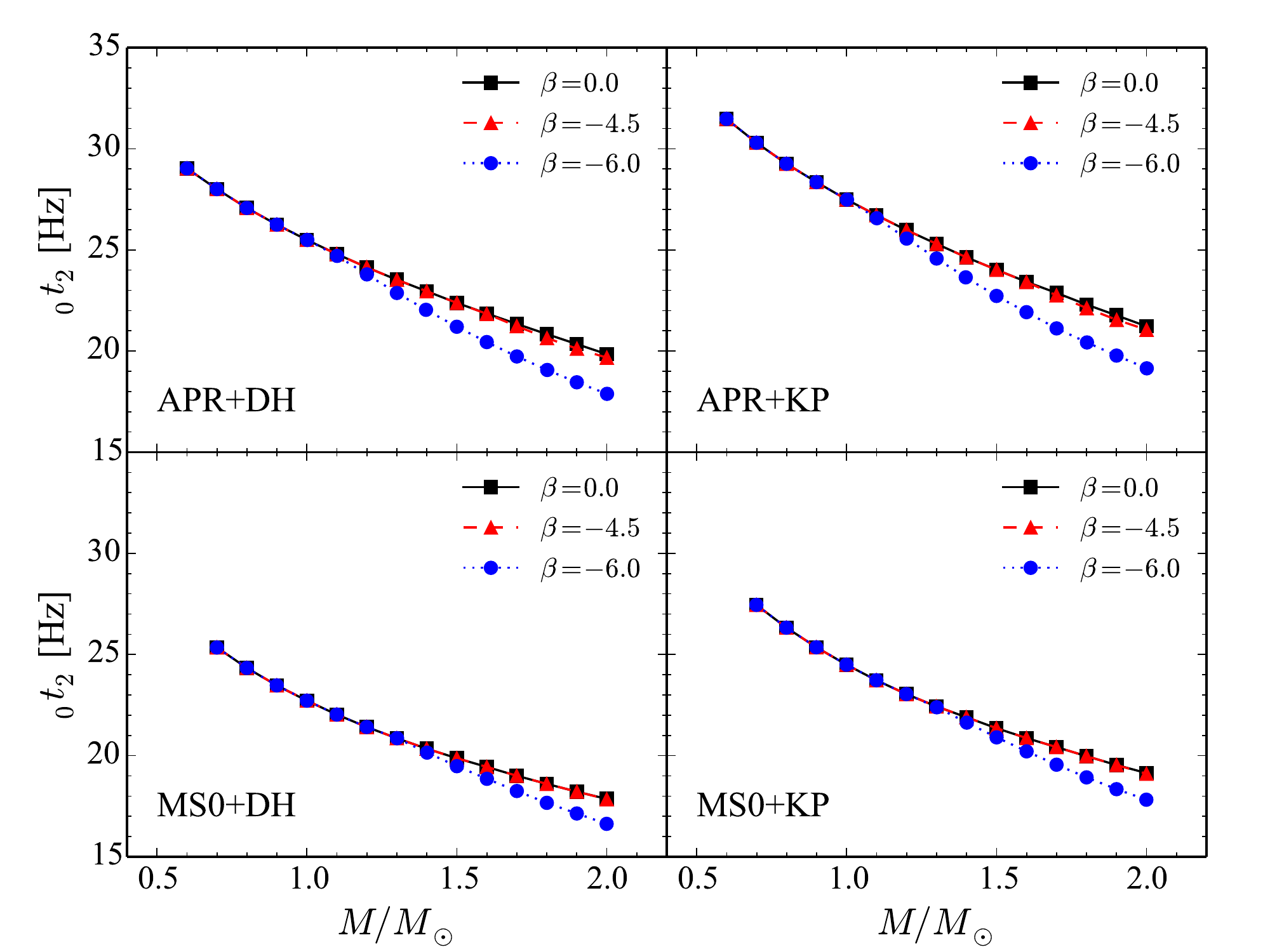}\\
\includegraphics[width=0.95\columnwidth]{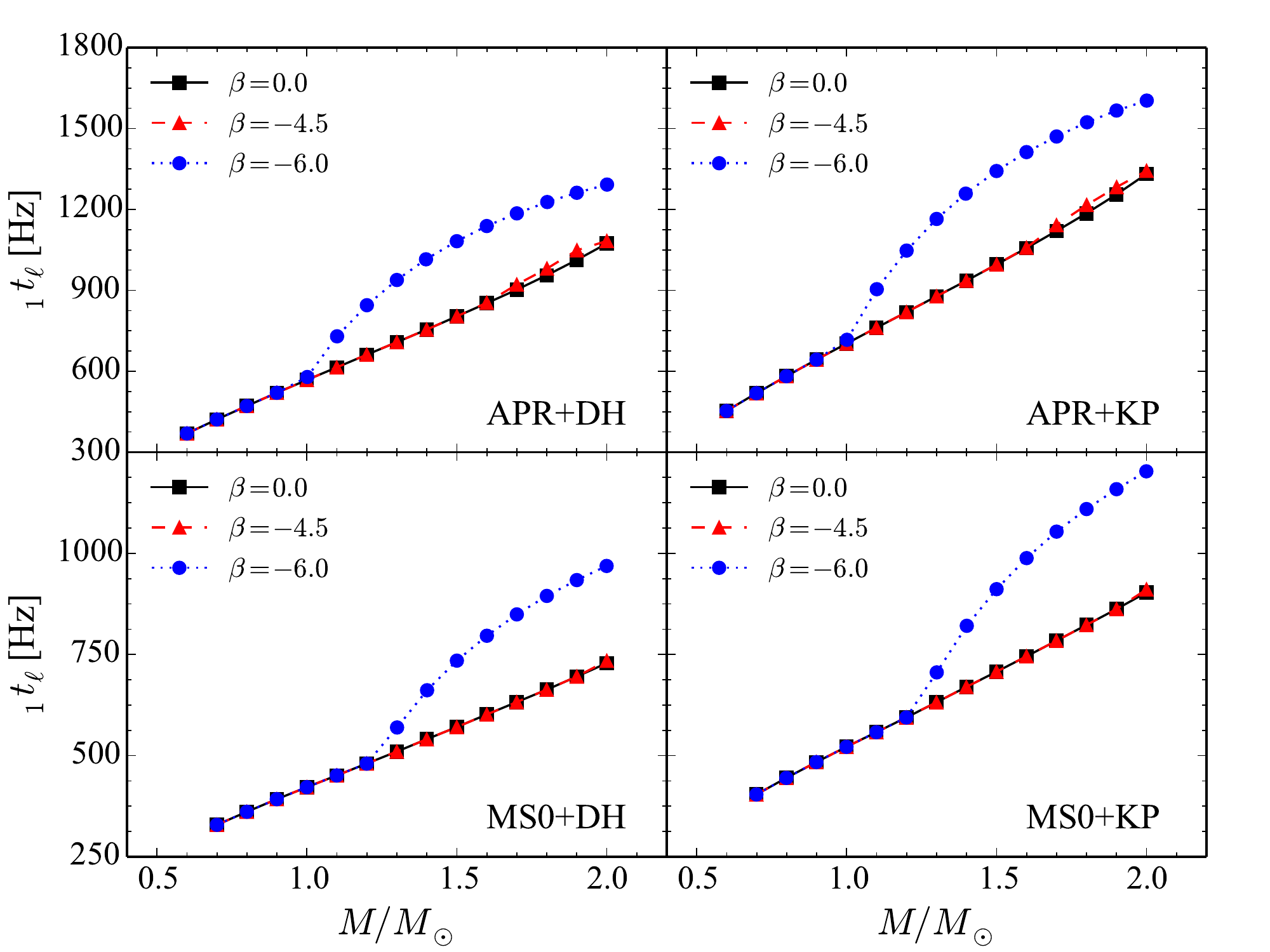}
\includegraphics[width=0.95\columnwidth]{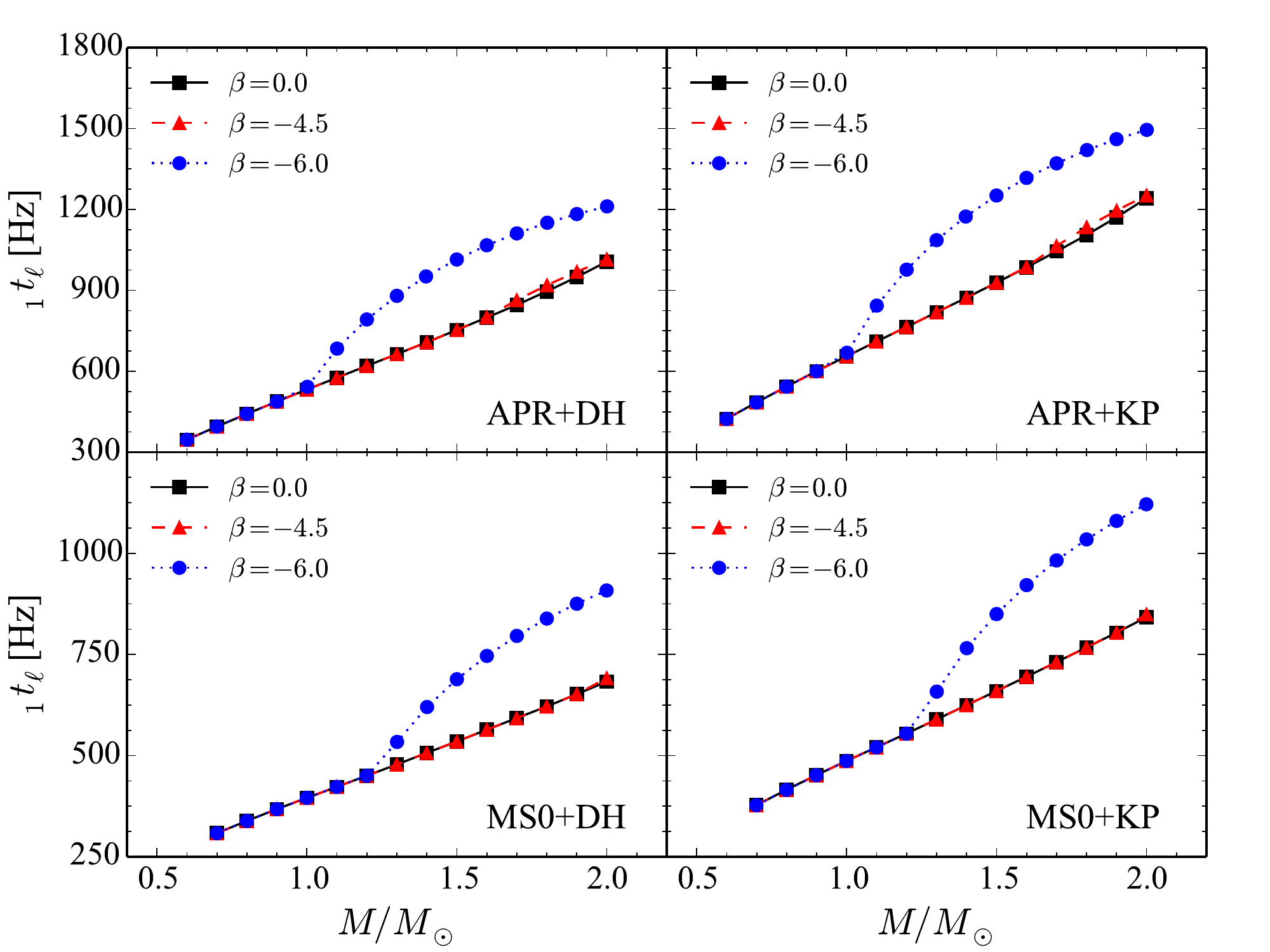}
\caption{(Color online) Frequencies of the torsional modes in
  scalar-tensor theory as a function of $M / M_{\odot}$. \textit{Top
    panels}: the fundamental torsional mode ${_0}t_{2}$ without ({\it
    left}) and with ({\it right}) electron screening. {\it Lower
    panels}: the first overtone ${_1}t_{\ell}$ without ({\it left})
  and with ({\it right}) electron screening.}
\label{freqs1}
\end{figure*}

\subsection{The shear modulus}\label{shear}

Torsional oscillations depend on the elastic properties of the solid
NS crust \cite{Chamel:2008ca}, characterized by the shear stress
tensor\footnote{Any deformation of an elastic medium can be decomposed
  into compressional and shear components. Matter in the NS crust is
  essentially incompressible, and this is why only a shear stress
  tensor is studied in the literature \cite{Chamel:2008ca,Ogata}.}. A
crucial element in describing the elastic properties of the NS crust
is the shear modulus $\muJ$. Assuming the NS crust to be a
body-centered cubic (bcc) lattice,
Ogata and Ichimaru \cite{Ogata} (see also \cite{Strohmayer}) showed
that the shear modulus in the limit of zero temperature can be
approximated as
\be
\tilde{\mu} = 0.1194 \,n\, \frac{(Ze)^2}{\tilde{a}},
\label{shear1}
\ee
where $n$ is the ion number density, $Ze$ the charge of the nuclei and
$\tilde{a}^3 = 3/(4\pi n)$ is the radius of the Wigner-Seitz cell
containing one nucleus. Although it is often assumed that the
electrons are uniformly distributed in the NS crust, one can also
calculate the correction to the shear modulus due a nonuniformity of
the electron density distribution, i.e. electron screening effects
\cite{Kobyakov:2013eta,Horowitz:2008xr}.  Kobyakov and Pethick
\cite{Kobyakov:2013eta} obtained the following electron screening
correction term to Eq.~(\ref{shear1}):
\be
\tilde{\mu} = 0.1194 \,n\, \left( 1 - 0.010\, Z^{2/3}\right) \frac{(Ze)^2}{\tilde{a}}.
\label{shear2}
\ee
For $Z=40$, electron screening can reduce the shear modulus by
$\approx 11.7 \%$. As discussed in \cite{Sotani:2014dua}, this reduces
the fundamental mode frequency ${_0}t_{2}$ by roughly $6\%$ in GR,
independently of whether we use EOS DH or KP.

In our calculations we consider both Eqs.~(\ref{shear1}) and
(\ref{shear2}) to see whether one would be able, in principle, to
distinguish modifications of the torsional oscillations spectrum due a
modified theory of gravity from microphysics effects (electron
screening being one of the simplest examples to investigate).

The impact of electron screening effects can be visualized by plotting
the shear velocity $\tilde{v}_s^2 = \muJ / (\rhoJ + \pJ)$ in the crust
region. Fig.~\ref{v_s} shows $\tilde{v}_s^2$ for NS models in GR and
in a scalar-tensor theory with $\beta = -6.0$, using both EOS DH and
KP, with and without electron screening effects. All NS models shown
in the figure have radius $R = 15.21$~km and mass $M = 2.046\,
M_{\odot}$. The (density-weighted) shear velocity
\be
\langle\tilde{v}_s\rangle = \frac{\int_{r_b}^{r_s} \rhoJ(r)\, \tilde{v}_s(r)\, r^2 \, dr}{\int_{r_b}^{r_s} \tilde{v}_s(r)\, r^2 \, dr}
\ee
is always close to $\approx 1\times 10^8$ cm/s, in remarkable
agreement with early estimates by Schumaker and Thorne
\cite{Schumaker} (see also \cite{1980ApJ...238..740H}).

\subsection{Numerical procedure}
To numerically integrate Eq.~(\ref{eq_y}) and obtain the frequencies
${_n} t_{\ell}$, it is convenient to introduce two new variables
$\widetilde{\cal Y}_1 (r)$ and $\widetilde{\cal Y}_2 (r)$, defined as
\begin{align}
\widetilde{\cal Y}_1(r) &\defeq r^{1-\ell}\, \widetilde{\cal Y}(r), \\
\widetilde{\cal Y}_2 (r) &\defeq \mueff \, e^{\Phi-\Lambda}\, r^{2-\ell}\, \widetilde{\cal Y}^{\prime}(r).
\end{align}
In terms of these variables, Eq.~(\ref{eq_y}) can be decomposed into a
system of two first-order coupled differential equations:
\begin{align}
\widetilde{\cal Y}^{\prime}_1 (r) &= -\frac{\ell-1}{r}\, \widetilde{\cal Y}_1(r) + \frac{e^{\Lambda-\Phi}}{\mueff\, r} \, \widetilde{\cal Y}_2(r) \label{y1}, \\
\widetilde{\cal Y}^{\prime}_2 (r) &=
- \frac{\ell+2}{r} \, \widetilde{\cal Y}_2 (r) - e^{\Phi+\Lambda}\left[ (\rhoJ+\pJ) \,r \,\bar{\omega}^2\, e^{-2\Phi} \right. \nonumber \\ &\left. - (\ell+2)(\ell-1) \frac{\mueff}{r}\right] \widetilde{\cal Y}_1(r). \label{y2}
\end{align}
The advantage of this approach is that it eliminates the necessity of
computing the derivative of the shear modulus $\muJ$, which is known
only in tabulated form.  In terms of $\widetilde{\cal Y}_2(r)$, the
zero-traction and zero-torque conditions translate into the
requirements that $\widetilde{\cal Y}_2 (r_b) = \widetilde{\cal Y}_2
(r_s)=0$. The same change of variables was used in
\cite{Sotani:2006at} in the context of magnetized stars (see also
\cite{Messios:2001br}).

Using Eqs.~(\ref{y1}) and (\ref{y2}) we can now find the frequencies
${_n} t_{\ell}$ by applying a shooting method (see
e.g.~\cite{Kokkotas:2000up}). Choosing $\widetilde{\cal Y}_1(r)$ to be
normalized to unity, and setting $\widetilde{\cal Y}_2(r)=0$ at the
stellar surface $r=R$, we integrate Eqs.~(\ref{dm})-(\ref{dp}),
(\ref{y1}) and (\ref{y2}) inwards for a trial value of $\omega$ until
we reach the crust basis at $r=r_b$, where we must have
$\widetilde{\cal Y}_2 (r_b)=0$. Depending on whether or not this
condition is satisfied, we adjust the trial value of $\omega$ until we
find $\widetilde{\cal Y}_2 (r_b)=0$ within a certain tolerance. In
this way the determination of $\omega$ becomes a root finding problem,
which can be solved using (for instance) the bisection method.

\section{The oscillation spectra}
\label{sec:spectra}

With our equilibrium NS models and our numerical framework to deal
with crustal perturbations, we are finally in a position to compute
and discuss the spectrum of torsional oscillation frequencies in
scalar-tensor theory. The spectrum depends quite sensitively on the
bulk properties of the star (mass $M$, radius $\widetilde{R}$, crust
thickness $\Delta \widetilde{R}$), on the choice of crustal EOS, and
on the scalar field profile in the crust region.

In Fig.~\ref{freqs1} we show the torsional oscillation frequencies for
the fundamental mode ${_0}t_{2}$ (top panels) and first overtone
${_1}t_{2}$ (bottom panels) as a function of the mass $M$ for NS
models with all possible combinations of core EOS (MS0, APR) and crust
EOS (DH, KP). We show results for three different values of $\beta$:
$\beta=0$ (GR), $\beta=-4.5$ (marginally excluded by binary pulsar
observations) and $\beta=-6$ (observationally excluded, but shown
nonetheless to maximize the effects of scalarization). By comparing
the left and right panels we can quantify the influence of electron
screening effects (everything else being the same): electron screening
typically lowers the oscillation spectra, in agreement with the
findings of Ref.~\cite{Sotani:2014dua}. For stellar models built using
EOS MS0 and for the conservative value $\beta = -4.5$, modifications
from GR occur at values of $M \simeq 2.0 \, M_{\odot}$, close to the
largest observed NS mass
\cite{Demorest:2010bx,Antoniadis:2013pzd}. Therefore from now on we
will focus on EOS APR.

\begin{figure}[h]
\includegraphics[width=\columnwidth]{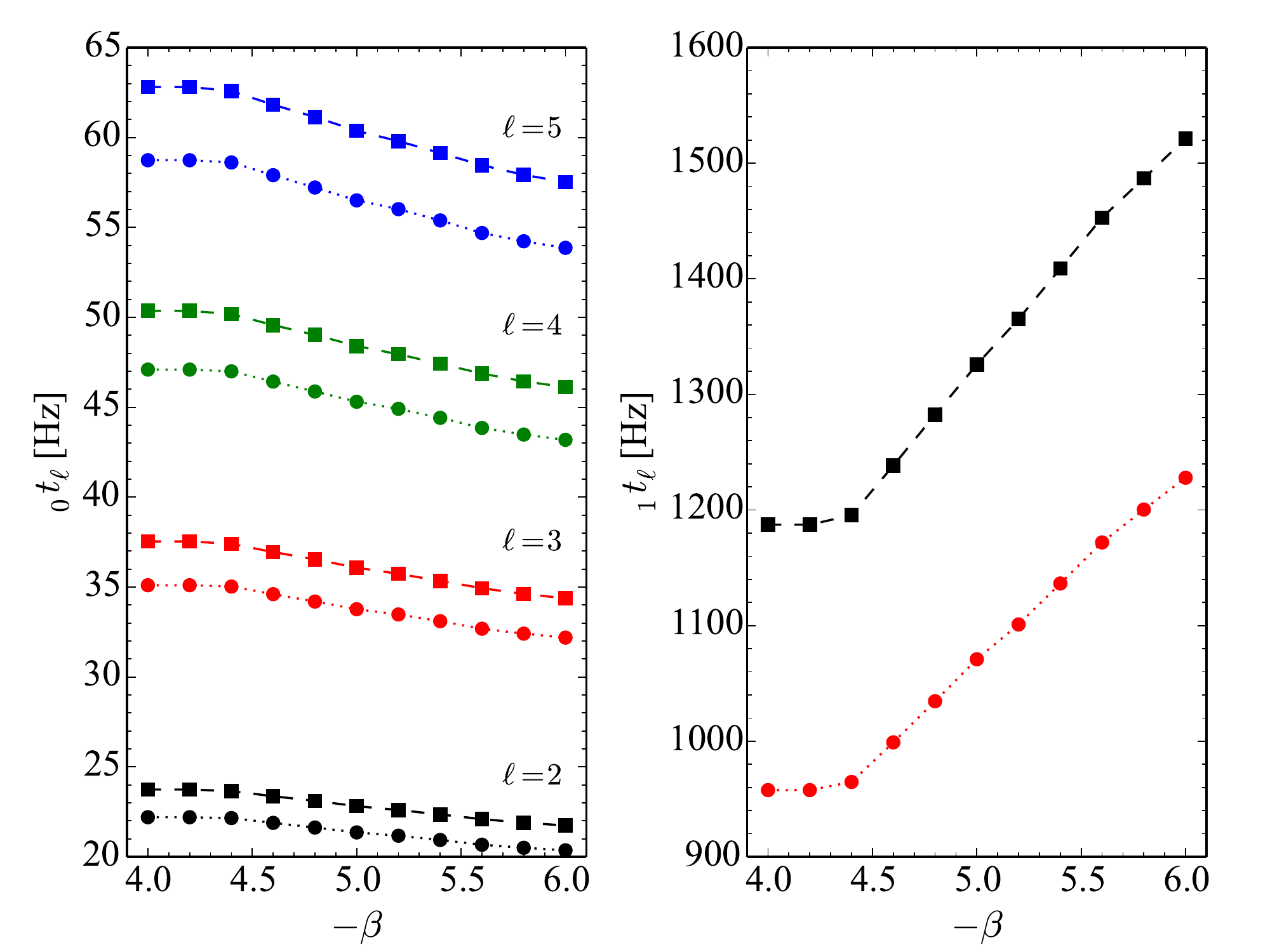}
\caption{(Color online) Frequencies of the torsional modes in
  scalar-tensor theory as a function of $\beta$ for stellar models
  with mass $M = 1.8 \, M_{\odot}$. Circles and dotted lines
  correspond to APR+DH; squares and dashed lines correspond to
  APR+KP. In the right panel we plot the mode frequencies
  ${_0}t_{\ell}$ for $\ell=2,3,4$ and 5. In the left panel we show the
  frequencies of the first overtone ${_1}t_{\ell}$.}
\label{freqs2}
\end{figure}

\begin{figure*}[thb]
\includegraphics[width=\columnwidth]{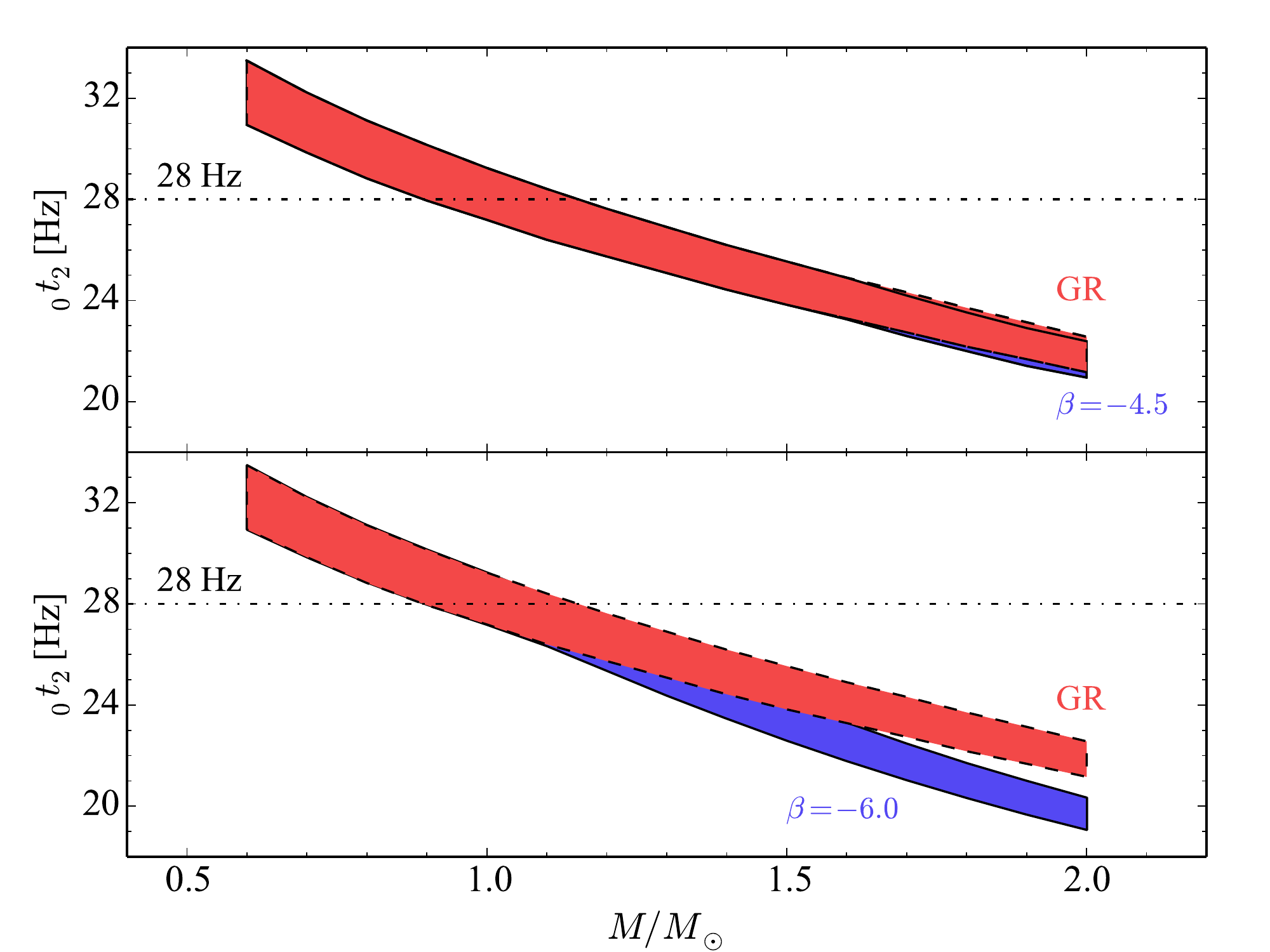}
\includegraphics[width=\columnwidth]{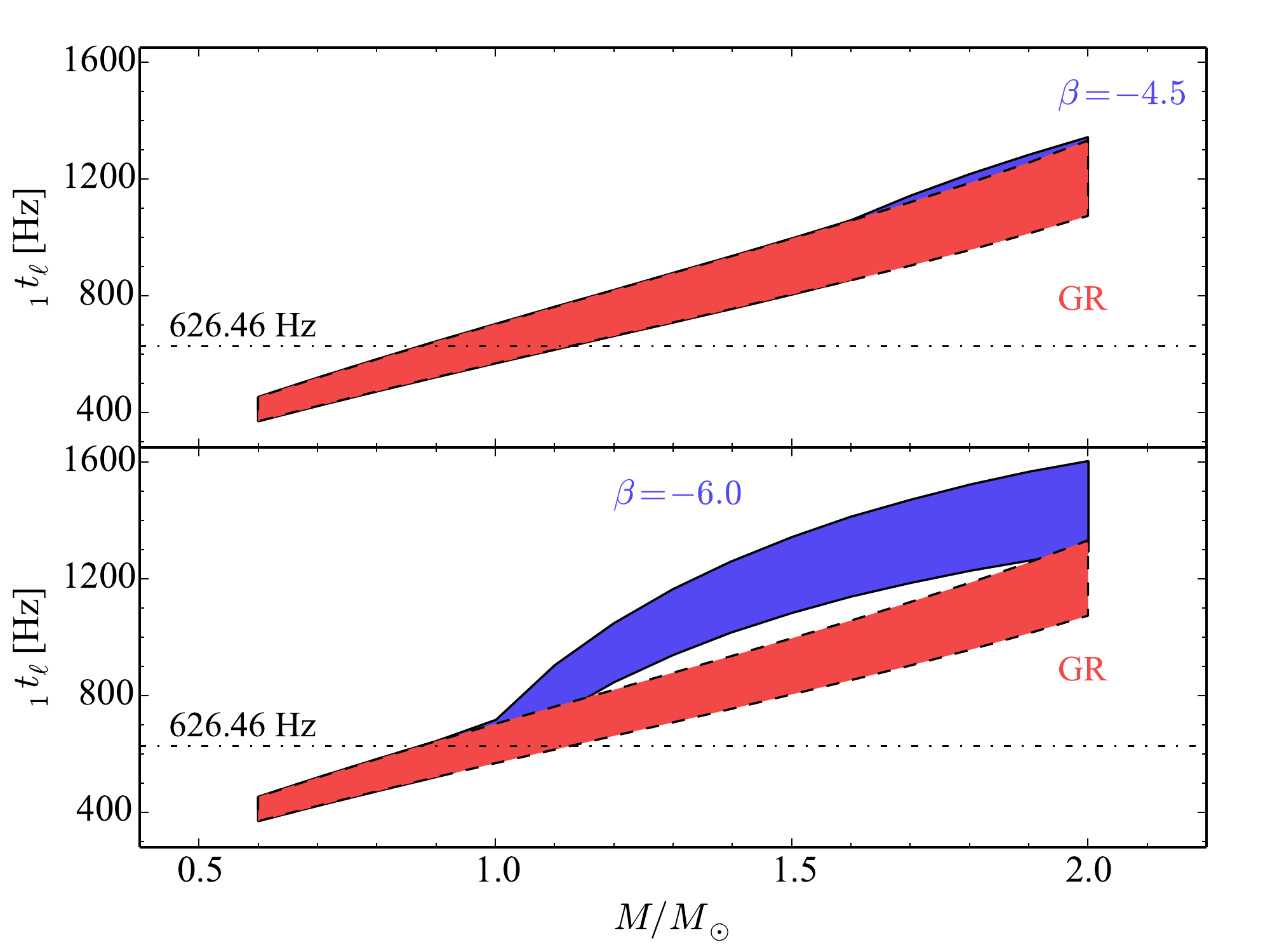} \\
\caption{(Color online) This plot compares modifications in torsional
  oscillation frequencies due to the underlying gravitational theory
  with crustal EOS uncertainties for models constructed using EOS APR
  in the core. Regions bounded by dashed lines correspond to
  oscillation frequencies in GR with different crustal EOSs; regions
  bounded by solid lines correspond to oscillation frequencies in
  scalar-tensor theory with different crustal EOSs. The degeneracy
  between modified gravity and crustal EOS is broken when the two
  regions do not overlap. Left panels refer to a scalar-tensor theory
  with $\beta = -4.5$, right panels to a theory with $\beta = -6.0$
  (a value already excluded by binary pulsar experiments
  \cite{Freire:2012mg}).}
\label{freqs3}
\end{figure*}

Notice that the first overtone is more sensitive to scalarization than
the fundamental mode.  This is confirmed in Fig.~\ref{freqs2}, where
we show the frequencies of the ${_0}t_{\ell}$ and ${_1}t_{\ell}$ modes
for a fixed stellar mass $M = 1.8\, M_{\odot}$ as a function of
$\beta$. Newtonian estimates \cite{1980ApJ...238..740H} (see also
\cite{Samuelsson:2006tt} for GR with similar conclusion), show that
the overtones scale roughly as $\approx n / \Delta \widetilde{R}$ and
are essentially independent of $\ell$, as long as $\ell$ is not much
larger than $n$. As shown by Eq.~(\ref{approx_st}) and in
Fig.~\ref{approx_crust}, scalarization decreases the crust thickness.
The shrinking crust thickness compensates for the reduced effective
shear modulus, and the net effect is an increase of the oscillation
frequencies. Notice also that in scalar-tensor theory the frequencies
of the fundamental torsional oscillation mode decrease as we decrease
$\beta$ (the opposite happens in tensor-vector-scalar theory
\cite{Sotani:2011rt}).

In Fig.~\ref{freqs3} we address the following question: are
uncertainties in the EOS small enough to allow for tests of the
underlying gravitational theory based on measurements of torsional
oscillation frequencies in QPOs? Unfortunately, the answer is in the
negative. Shaded regions in the plot are bounded by the values of the
torsional oscillation frequencies computed using EOS DH and KP for the
crust. One region (bounded by dashed lines) corresponds to GR, while
the other (solid lines) to scalar-tensor theory. These regions are
meant to roughly quantify the EOS uncertainty {\em within each
  theory.}
Horizontal lines in the left panels mark the QPO frequency of $28$~Hz
observed in SGR 1900+14 \cite{Strohmayer:2005ks}, and identified with
the ${_0} t_{2}$ mode. The plots show that for a theory parameter
$\beta = -4.5$ (marginally ruled out by binary pulsar observations
\cite{Freire:2012mg}) the predictions of GR and scalar-tensory theory
are indistinguishable within uncertainties in the crustal EOS. The
bottom-left panel shows that, in principle, a scalar-tensor theory
with $\beta = -6.0$ could be distinguished from GR if we were to
observe QPOs with frequencies smaller than 24 Hz in magnetars with $M
\gtrsim 1.6\, M_{\odot}$. However, such a large value of $\beta$ is
already excluded by binary pulsar experiments.
The right panel carries out a similar analysis for the first overtone
${_1} t_{\ell}$. The horizontal line indicates the QPO frequency of
$626.46 \pm 0.02$~Hz detected in SGR 1806-20 \cite{Strohmayer:2006py},
and identified with the first overtone ${_1} t_{\ell}$. The
conclusions are similar: for $\beta = -4.5$, the predictions of GR and
scalar-tensory theory are indistinguishable within uncertainties in
the crustal EOS.

Let us now focus on the fundamental mode ${_0} t_{2}$, which has been
identified with QPOs in both SGR 1900+14 ($28 \pm 0.5$ Hz)
\cite{Strohmayer:2005ks} and SGR 1806-20 ($30.4 \pm 0.3$ Hz)
\cite{Strohmayer:2006py}. To quantify the relative effect of
scalarization and electron screening, assuming the crustal EOS to be
known, we introduce the ratio
\be\label{eta_def}
\eta \defeq \frac{ \vert {_0}t_{2}[\text{ST}] - {_0}t_{2}[\text{GR}]\vert }{\vert {_0}\bar{t}_{2}[\text{GR}] - {_0}t_{2}[\text{GR}] \vert},
\ee
where ${_0}t_{2}[\text{GR}]$ (${_0}t_{2}[\text{ST}] $) is the
fundamental mode frequency in GR (scalar-tensor theory) ignoring
electron screening, and ${_0}\bar{t}_{2}[\text{GR}]$ is the
corresponding frequency in GR computed by taking into account electron
screening. Electron screening has a larger impact than scalarization
whenever $\eta<1$.

\begin{figure}
\includegraphics[width=\columnwidth]{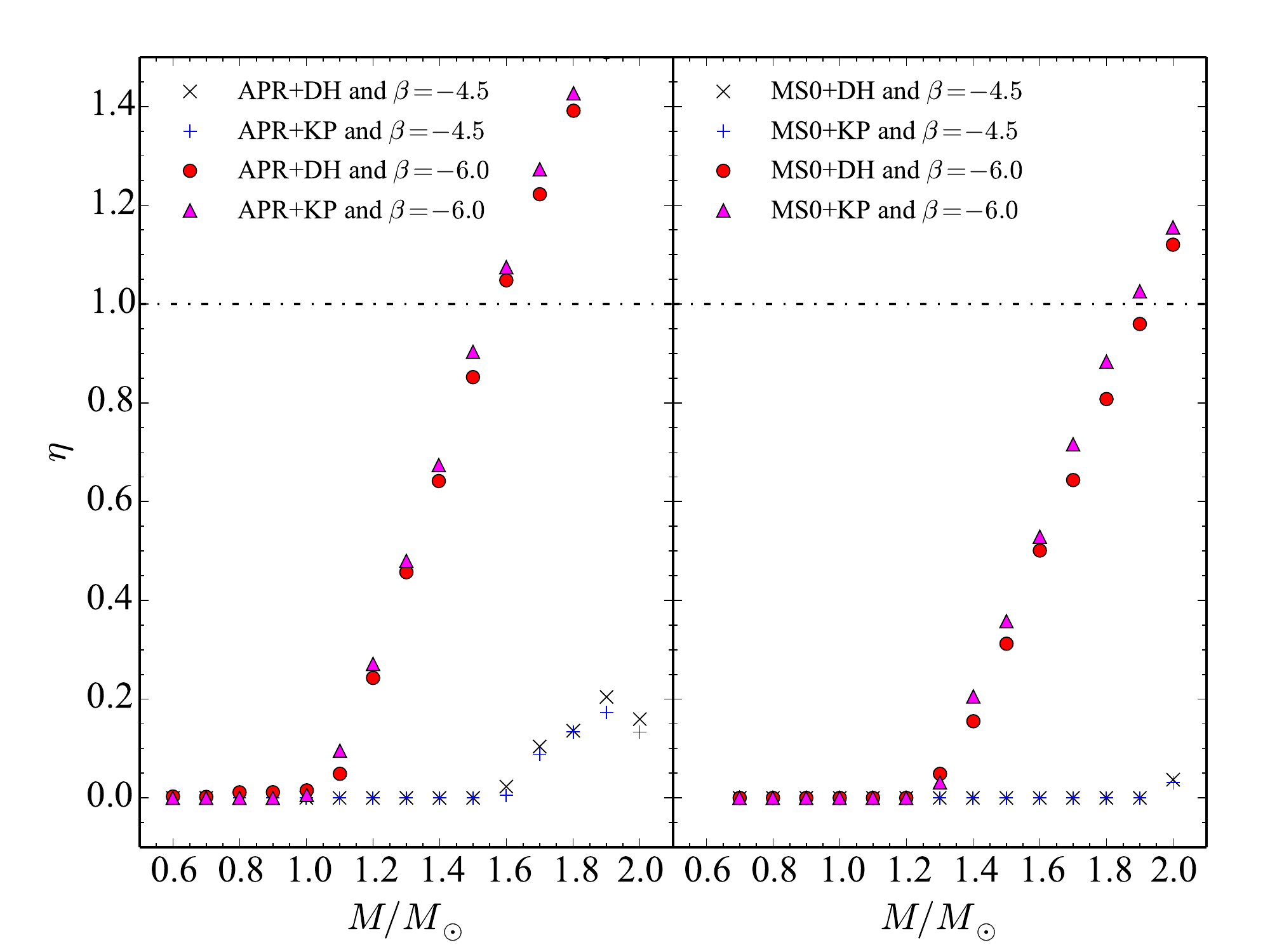}
\caption{(Color online) The ratio $\eta$ defined in
  Eq.~(\ref{eta_def}) for all stellar models considered in this
  work. Values of $\eta > 1$ mean that the effect of scalarization is
  larger than that of electron screening. This would only be possible
  for values of $\beta$ that are already ruled out by binary pulsar
  experiments.}
\label{freqs5}
\end{figure}

In Fig.~\ref{freqs5} we show $\eta$ as a function of the mass $M$ for
all combinations of core and crust EOS considered in this work. The
punchline of this plot is consistent with our previous findings: the
effect of electron screening is always dominant over scalarization for
values of $\beta$ that are compatible with current binary pulsar
experiments. Unrealistically large values of $\beta$ (e.g.,
$\beta=-6$) would be needed to constrain scalar-tensor theories via
torsional oscillation frequencies.

%

\section{Conclusions}
\label{sec:conclusions}

We studied torsional oscillations in NS crusts in scalar-tensor
theories of gravity allowing for spontaneous scalarization. Working in
the Cowling approximation, we showed that the ``master equation''
governing torsional oscillations -- our Eq.~(\ref{eq_y_2}) -- has the
same form as in GR \cite{Schumaker} if we introduce an effective shear
modulus $\muJ_{\text{eff}}$, an effective wave velocity $\veleff$ and
a rescaled frequency $\bar{\omega}$. In general, a smaller effective
shear modulus reduces the oscillation frequencies. However we showed
both analytically and numerically that the NS crust becomes thinner
under scalarization, and a thinner crust tends to increase the
overtone frequencies. Our numerical calculations show that the reduced
shear modulus is the dominant effect for the fundamental mode, while
the change in crust thickness is dominant for the first overtone.

We found that the dominant torsional oscillation frequencies in
scalar-tensor theory are essentially indistinguishable from those in
GR for all values of $\beta \geq -4.5$ that are still allowed by
binary pulsar observations. One of the simplest microphysics effects
that might affect the torsional oscillation frequencies, namely
electron screening \cite{Sotani:2014dua}, has a much more important
effect on torsional oscillation frequencies than scalarization. More
noticeable deviations from GR would occur for (say) $\beta = -6.0$,
but such large values of $\beta$ are already ruled out by
binary-pulsar observations \cite{Freire:2012mg}. We expect
scalarization to be subdominant when compared to other uncertainties
in the microphysics, such as nonuniform nuclear structures (pastas)
\cite{2011MNRAS.417L..70S} and superfluidity of dripped neutrons
\cite{Sotani:2012xd}.

Given the similarities between torsional oscillation frequencies in GR
and scalar-tensor theory, we can conjecture that the inclusion of slow
rotation in our model will result in torsional modes growing due to
the Chandrasekhar-Friedman-Schutz (CFS) instability
\cite{Vavoulidis:2007ui}. The inclusion of slow rotation adds an extra
term proportional to the frame dragging function $\varpi$
(cf.~\cite{Hartle:1967he}) in the perturbation equation
(\ref{eq_y_2}). Previous studies of slowly rotating NSs in
scalar-tensor theory \cite{Sotani:2012eb} showed that scalarization
affects $\varpi$, and therefore it will affect torsional modes for
rotating stars.

One important omission in our study is the effect of magnetic fields, a
crucial ingredient for realistic comparisons with QPO observations in
magnetars.  Very few works have studied NSs with magnetic fields in
alternative theories of gravity (see e.g.~\cite{Hakimov:2013zoa}).
Couplings between the scalar field and magnetic fields may produce
larger deviations of the torsional oscillations frequencies with
respect to GR. This is an interesting topic for future study.

\section*{Acknowledgements}

We are grateful to P.~Pani for useful discussions and for validating
some of our results. H.~O.~S., E.~B. and M.~H. were supported by NSF
CAREER Grant No.~PHY-1055103. H.~S. was supported by Grant-in-Aid for
Young Scientists (B) through JSPS Grant No.~26800133.

\appendix
\section{Derivation of Eq.~(\ref{approx_st})}
\label{app1}

In this Appendix we present the derivation of
Eq.~(\ref{approx_st}). Making use of Eq.~(\ref{dphi}), we rewrite
Eq.~(\ref{dp}) as
\begin{align}
\frac{d\pJ}{dr} &= - (\rhoJ + \pJ) \left[ 4 \pi A^4(\vp) \frac{r^2 \pJ}{r-2m} + \frac{1}{2}r\psi^2  \right. \nonumber \\
& \left. + \frac{m}{r(r-2m)} + \alpha(\vp)\psi \right].
\label{dp2}
\end{align}
Let us assume that the following approximations hold true in the NS
crust: (i) $m_s \approx M$, and therefore $e^{-2\Lambda} = 1 - 2
M/r_s$; (ii) the pressure $\pJ$ is negligible in comparison to $\rhoJ$
\cite{Samuelsson:2006tt}; (iii) $\varphi \approx \varphi_s$ and $\psi
\approx \psi_s$; (iv) $A(\varphi) \approx 1$. We also assume that the
EOS has the polytropic form $\rhoJ = k \pJ^{1/\Gamma}$, where $k$ and
$\Gamma$ are constants.
Then Eq.~(\ref{dp2}) becomes
\be
\frac{d\pJ}{dr} \approx - \pJ \, e^{2 \Lambda}\,\frac{M}{r} - \rhoJ \left[ \frac{1}{2} r \psi_s^2 + \alpha(\varphi_s) \psi_s \right],
\ee
where $\alpha(\varphi_s) = \beta \varphi_s$. Integrating this equation from $r=r_b$ to $r=r_s$ and imposing $\pJ(r_s)=0$ we obtain
\begin{align}
0 &= \sigma + M e^{2\Lambda} \left( \frac{1}{r_s} - \frac{1}{r_b} \right) - \psi_s^2 (r_s^2 - r_b^2) \nonumber \\
&- \alpha(\varphi_s) \psi_s (r_s-r_b),
\label{approx2}
\end{align}
where we have defined $\sigma \defeq \xi \pJ_b / \rhoJ_b$ and $\xi
\defeq \Gamma / (\Gamma-1)$ (recall that the subscript $b$ denotes
quantities evaluated at the crust basis).

We now make the additional assumption that $\psi_s^2 (r_s^2 - r_b^2)$
is negligible compared to $\alpha(\varphi_s) \psi_s (r_s-r_b)$. We
have verified this assumption by explicitly evaluating these two terms
for different stellar models: typically $\alpha(\varphi_s) \psi_s
(r_s-r_b)$ is larger than $\psi_s^2 (r_s^2 - r_b^2)$ by at least a
factor 10.

Rewriting Eq.~(\ref{approx2}) in terms of ${\cal R}$ we obtain the
quadratic equation
\be
0 = \frac{\beta \xi}{\sigma}\, {\cal R}^2 - \left[ 1+\frac{1}{\sigma} \left( {\cal C}\, e^{2\Lambda} + \beta \zeta \right) \right] {\cal R} +1,
\label{quadratic}
\ee
where we introduced $\zeta = \zeta({\cal C}) \defeq \varphi_s \,
\psi_s \, r_s$, which must be obtained by interpolation, given a
family of stellar models, as a function of ${\cal C}$. Choosing the
solution of Eq.~(\ref{quadratic}) that reduces to the GR result
(\ref{approx_gr}) when $\beta \rightarrow 0$ and defining ${\cal F}
\defeq 1 + \left(\,{\cal C} e^{2\Lambda} + \beta \zeta \,
\right)/\sigma$, we finally obtain Eq.~(\ref{approx_st}).

\section{Equivalence of the perturbation equations in Einstein and Jordan frames}
\label{app2}

Here we show that the perturbation equation (\ref{pert_j}) could also
be obtained by starting with the energy-momentum conservation law in
the Einstein frame,
\be
\nabla_{\ast_{\mu}} T_{\ast}^{\mu\nu} - \alpha(\vp) T_{\ast} \nabla_{\ast}^{\nu}\vp = 0. \nonumber
\ee
For odd (axial) perturbations in the Cowling approximation, the
perturbed Einstein-frame energy-momentum tensor $\delta T_{\ast
  \mu\nu}$ satisfies
\be
\pa_{\mu}\delta T^{\mu}_{\ast \nu} + \Gamma^{\mu}_{\ast\sigma\mu} \delta T^{\sigma}_{\ast\mu} - \Gamma^{\sigma}_{\ast\nu\mu} \delta T^{\mu}_{\ast\sigma} - \alpha(\vp) \delta T_{\ast} \nabla_{\ast}^{\nu}\vp = 0.
\label{b1}
\ee
Using the relation $T^{\mu}_{\ast\nu} = A^{4}(\vp)
\widetilde{T}\indices{^\mu_\nu}$ -- which implies $\delta
T^{\mu}_{\ast\nu} = A^{4}(\vp) \delta \widetilde{T}\indices{^\mu_\nu}$
-- and the trace relation $T_{\ast} = A^4(\vp) \widetilde{T}$, we
obtain upon substitution into Eq.~(\ref{b1}) that
\begin{align}
& 4 A^{3}(\vp) \frac{A(\vp)}{d\vp}\pa_{\mu} \delta \widetilde{T}\indices{^\mu_\nu} + A^{4}(\vp) \left[ \pa_\mu \delta\widetilde{T}\indices{^\mu_\nu} \right.  \nonumber \\ 
&+ \left. \Gamma^{\mu}_{\ast\sigma\mu}\delta\widetilde{T}\indices{^\sigma_\nu} - \Gamma^{\sigma}_{\ast\nu\mu} \delta \widetilde{T}\indices{^\mu_\sigma} - \alpha(\vp) \pa_{\nu}\vp\, \delta \widetilde{T} \right]=0.
\end{align}
Dividing by $A^4(\vp)$ we recover Eq.~(\ref{pert_j}).


%

\end{document}